\documentclass[11pt,a4paper]{article}
\usepackage[T1]{fontenc}
\usepackage{lmodern,amsmath,amssymb,amsthm,mathtools,bm,booktabs,microtype,tikz}
\usepackage[margin=28mm]{geometry}
\definecolor{ink}{HTML}{234F66}
\definecolor{accent}{HTML}{8B4931}
\usepackage[colorlinks=true,linkcolor=ink,citecolor=ink,urlcolor=ink]{hyperref}
\hypersetup{pdftitle={Same Pieces, Different Particles: Identity and Conversion in Supersymmetric Matrix Scattering},pdfauthor={Mai Qi}}
\numberwithin{equation}{section}
\newtheorem{theorem}{Theorem}[section]

\newtheorem{corollary}[theorem]{Corollary}
\newcommand{\Tr}{\operatorname{Tr}}
\newcommand{\Pf}{\operatorname{Pf}}
\newcommand{\ad}{\operatorname{ad}}
\newcommand{\Spin}{\operatorname{Spin}}
\newcommand{\Hil}{\mathcal H}
\newcommand{\ac}{\mathrm{ac}}
\newcommand{\pp}{\mathrm{pp}}
\newcommand{\scs}{\mathrm{sc}}
\newcommand{\R}{\mathbb R}
\newcommand{\C}{\mathbb C}
\newcommand{\arref}[1]{\href{https://arxiv.org/abs/#1}{arXiv:#1}}
\title{Same Pieces, Different Particles:\\
Identity and Conversion in Supersymmetric Matrix Scattering}
\author{Mai Qi\textsuperscript{1,2}\\[0.5em]
  {\small\textsuperscript{1}\,University of Cambridge, Cambridge, United Kingdom}\\
  {\small\textsuperscript{2}\,University of California, Santa Barbara, California, USA}\\[0.3em]
  {\small\href{mailto:maiqi@ucsb.edu}{\nolinkurl{maiqi@ucsb.edu}}}}
\date{}
\begin{document}
\maketitle

\begin{abstract}
A bound cluster of color matrices can emerge from a collision with its internal state intact, sharing its rest energy and relative momentum with a differently embedded fragment. We determine which bound clusters exhaust the scattering states of pure Yang--Mills quantum mechanics with four or eight real supercharges for each fixed compact, simply connected simple gauge group. To establish completeness we follow interacting subsystems through successive separations and control the geometric phase of their occupied fermions. In four-supercharge \(SU(3)\), the four-fermion continuum begins at the positive ground energy of a bosonic dimer. In four-supercharge \(\Spin(2n)\), for every even \(n\ge4\), the two gauge-inequivalent spinor embeddings of a bosonic \(SU(n)\) ground fragment give opposite asymptotic Pfaffian readings. At fermion number \(n(n-1)\) and total spin zero, a collision converts one embedding into the other with nonzero probability at almost every energy between the fragment's ground and first excited levels. We analytically continue a coupling that is strictly positive at a negative energy and recover the transmission amplitude from its boundary value. For a doubly raised incoming state, the division of energy between rest and recoil fixes the relative weights of three outgoing polarizations with zero, one, or two extra fermions on the fragment.
\end{abstract}

\clearpage
\tableofcontents
\clearpage

\section{Introduction}

Suppose a collision in matrix quantum mechanics ejects a bound fragment. A distant observer measures its rest energy and the momentum of its center. In the four-supercharge \(\Spin(12)\) model these data agree for two ways of embedding the same bosonic \(SU(6)\) ground state in the color algebra. The embeddings are gauge inequivalent, and an observer who also measures the Pfaffian of the separating matrices can tell them apart. We will show that a fragment prepared with one Pfaffian sign has positive probability of emerging with the other.

The first ingredient of this experiment is the cancellation of the transverse oscillator energy along a commuting valley. De Wit, L\"uscher and Nicolai used it to construct the continuum of the supersymmetric matrix Hamiltonian~\cite{dWLN}. An internally bound subsystem can supply a branch of that continuum~\cite{Michishita}, and at fixed rank in the sixteen-supercharge \(SU(N)\) theory the channels built from such bound states exhaust the continuous space~\cite{Qi}. Unbroken gauge algebras also figure in the low-energy spectrum of Yang--Mills theory on a torus~\cite{HW}, the effective motion of regular Cartan configurations~\cite{Smilga}, and the continuum contribution to a supersymmetric index~\cite{LeeYi}. The index takes a signed trace over states, while a scattering experiment follows a particular incoming wave and measures how much of its norm reaches a chosen outgoing channel.

The number of occupied fermions can already decide which fragment is lightest. In four-supercharge \(SU(3)\), four fermions suffice to cancel the oscillator energy between a pair of colors and a singleton. All four must occupy those connecting modes, leaving the pair with the Hamiltonian of purely bosonic \(SU(2)\) matrix mechanics. Although its classical potential vanishes on commuting configurations, its lowest quantum energy \(e_b\) is positive. We identify \(e_b\) as the exact onset of the four-fermion continuum, while six fermions allow three singletons to separate at arbitrarily small energy. The distinction bears directly on finite-basis studies of the same three-color Hamiltonian~\cite{Ambrozinski}.

A general separating configuration leaves an unbroken gauge algebra. Its center carries the free relative motion, while each simple factor can support a fluctuating bound state. If one factor subsequently breaks up, the same final particles admit a direct separation and a separation in two stages. In the eight-supercharge theory the occupied fermions can acquire a relative Berry phase along these two separation paths. Berry phases also occur in the excited modes of a separated pair~\cite{PST}. For an interacting subsystem the internal commutator that generates the relative phase also enters the internal potential, which bounds the phase error even when the fragment extends far along a nearly commuting valley. Comparing the kinetic energies of the two waves also requires differentiating the change of coordinates between the two descriptions. A stretched coordinate has nonzero charge under the separating center, so its first-order displacement cannot move a neutral internal coordinate. Together these estimates allow the construction of Ref.~\cite{Qi} to retain an interacting subsystem in the reduced theories.

Knowing all channels still leaves their transition probabilities undetermined. For four-supercharge \(\Spin(2n)\) with even \(n\ge4\), each of the two spinor embeddings supports one radial wave in its first internal band at \(F=n(n-1)\) and total spin zero. The issue is whether a collision excites precisely the wave that can escape down the second valley. We follow a solution normalized in one valley as its energy passes from below the spectrum into the band. At negative energy, quaternionic pairing of the occupied fermions fixes the sign of the coupling between the two valleys. At positive energy, a test with the outgoing radial wave identifies that same coupling with the physical transmission amplitude times two nonzero normalization factors. Boundary uniqueness then implies conversion at almost every energy in the band, even if other exits are present. Symmetry fixes transmission into two transverse valleys in a toy supermembrane model~\cite{BRY}. For scalar Schr\"odinger operators on spaces with distinct geometric ends, Ito and Skibsted prove injectivity of transmission under their geometric hypotheses~\cite{Ito}. Here all the valleys belong to a single Cartesian space, and a finite-distance overlap can involve modes that carry no flux to infinity.

At \(n=4\), triality permutes three embeddings and the first-band scattering matrix has only two independent eigenphases. For a fragment incident in one channel, the probabilities of emerging in either of the other two are equal and depend only on the difference of these eigenphases. At least one ninth of the incident flux returns to the original channel. At every even rank a second measurement is possible after raising the incoming state by two supercharges. Resolving the two extra fermions between the internal state and the relative motion gives three probabilities whose ratios are fixed by the division of energy between rest and recoil.

Section~\ref{sec:dimer} locates the dimer threshold. Section~\ref{sec:channels} identifies the complete channel space and computes the phase for a fragment that separates in two stages. Section~\ref{sec:triality} constructs the Pfaffian measurement, proves conversion for the spinor embeddings and derives the triality and polarization probabilities. Appendices~\ref{app:channels} and~\ref{app:transmission} give the estimates for successive breakups and the continuation from negative energy to the physical amplitude, respectively.

\section{A dimer at the bottom of a continuum}\label{sec:dimer}

\subsection{The relative Hamiltonian and the price of separation}
Let \(\mathfrak g\) be the real compact Lie algebra of the gauge group with a positive invariant inner product. In orthonormal color coordinates the \(d\) matrices \(X_\mu\) have flat kinetic energy. We take \(d=3,5\), with respectively four and eight real supercharges, positive coupling \(g\) and \(\nu=d-1\) complex fermion modes per real generator. In the number-preserving polarization the Hamiltonian and the physical Hilbert space are
\begin{align}
 H_G&=-\tfrac12\Delta_X+\tfrac{g^2}{2}\sum_{\mu<\lambda}
    \|[X_\mu,X_\lambda]\|^2+
    \mathrm d\Gamma\!\left(g\sum_\mu\Gamma_\mu\otimes i\ad X_\mu\right),\label{eq:H}\\
 \Hil_G&=\big[L^2(\mathfrak g^d)\otimes
      \Lambda^\bullet(\C^\nu\otimes\mathfrak g_\C)\big]^G .\nonumber
\end{align}
The Hermitian spin matrices obey \(\{\Gamma_\mu,\Gamma_\lambda\}=2\delta_{\mu\lambda}\), the adjoint action is real skew in an orthonormal compact color basis, and \(\mathrm d\Gamma\) sums the one-body mass over the occupied modes. Exterior degree gives the conserved fermion number \(F\). In three spatial dimensions \(J\) denotes total angular momentum, including the fermion spin. All subsystem metrics are induced by the parent and the freely translating overall \(U(1)\) has been removed. Writing \(X=g^{-1/3}\widetilde X\) factors \(g^{2/3}\) out of \(H_G\). We retain \(g\) in the confinement estimate below and use \(g=1\) in the channel and scattering calculations.

On gauge singlets the supercharge anticommutator gives \(H_G\ge0\). Smooth compactly supported singlets form an operator core. With \(\langle X\rangle=(1+|X|^2)^{1/2}\), the matrix potential is bounded below by \(-C\langle X\rangle\), and the metric \((1+C\langle X\rangle)^{-1}dX^2\) is complete. The weight has Lipschitz inverse square root, and the smooth potential gives the local domain hypothesis by elliptic regularity. The essential self-adjointness criterion of Braverman, Milatovic and Shubin~\cite[Corollary~2.9]{BMS} applies to the Cartesian operator as in Ref.~\cite[Sec.~2.1]{Qi}. Compact gauge averaging commutes with \(H_G\) and preserves smooth compact support. The same argument with rotation averaging gives a core in the scalar sector.

In four-supercharge \(SU(3)\), an unordered pair of separated colors gives two opposite root spaces. On each the leading mass is a Pauli matrix contracted with the separation vector, with one positive and one negative eigenvalue. Filling the negative state in both orientations gives a neutral two-fermion state whose energy cancels the transverse bosonic zero-point energy. Leaving either state empty costs energy proportional to the separation. A pair and a singleton stretch two root pairs, while three singletons stretch all three. With four fermions only the former can reach infinity at finite energy, and the dimer and relative center must both be in degree zero.

The surviving dimer has only its bosonic Hamiltonian \(H_b\). Along a commuting configuration its classical potential vanishes, but displacing one of the matrices creates transverse oscillators in the others. The bound will also apply to the larger bosonic fragments in Section~\ref{sec:triality}, so consider any compact simple color algebra \(\mathfrak s\) with its inherited metric. Reserve \(-\Delta/4\) of the internal kinetic energy and distribute the rest, together with the potential, over the six ordered pairs \(i\ne j\). At fixed \(X_j\) the operator for one pair is
\[
 -\tfrac18\Delta_{X_i}+\tfrac{g^2}{4}\|[X_j,X_i]\|^2 .
\]
A singular value \(\lambda\) of \(\ad X_j\) gives a one-dimensional oscillator of ground energy \(g\lambda/(4\sqrt2)\), with zero lower bound in a null direction. Each \(j\) occurs twice, so the six oscillator inequalities give the form bound
\begin{equation}
 H_b\ \geq\ -\tfrac14\Delta+
       \frac{g}{2\sqrt2}\sum_{j=1}^3\Tr_{\mathfrak s_\C}|i\ad X_j|.
 \label{eq:bosbound}
\end{equation}
For \(\mathfrak{su}(2)\) in the orthonormal basis with \([T_a,T_b]=\epsilon_{abc}T_c\), the color trace is \(2|X_j|\). More generally, simplicity makes the trace strictly positive on the unit sphere of \(\mathfrak s\), hence bounded below by \(c|X_j|\). Inequality~\eqref{eq:bosbound} confines a wave of bounded energy to a bounded region up to arbitrarily small tails. Local compactness gives compact resolvent, and a zero-energy state would have to be a square-integrable constant. The scalar heat kernel is positivity improving, so the unique normalized positive ground state is invariant under spatial rotations and every isometric color automorphism. Bosonic confinement and the scalar ground state also enter L\"uscher's analysis of Yang--Mills theory at small volume~\cite{Luscher}.

For the dimer, write \(e_b\) for this ground energy in the metric induced by \(SU(3)\). More explicitly, if \([T_a,T_b]=\kappa\epsilon_{abc}T_c\) in an orthonormal basis of the \(\mathfrak{su}(2)\) subsystem, then \(e_b=(g\kappa)^{2/3}\varepsilon_b\), where \(\varepsilon_b\) is the dimensionless bosonic \(SU(2)\) ground energy at unit structure constant and coupling. The isolated bosonic level computed in finite-basis studies~\cite{Wosiek} thus supplies the rest energy of an escaping dimer in the three-color system.

\subsection{Locating the threshold from both sides}
Place the dimer in this ground state and send its mass-normalized relative center \(z\in\R^3\) out to radius \(L\). At fixed wave number \(k\geq0\), a shell of width \(\ell=\sqrt L\) recedes from the origin while its spread in kinetic energy tends to zero. With \(r=|z|\) and a smooth unit envelope \(\chi\) supported in \((-1,1)\), the spherical wave
\begin{equation}
 v_L(z)=\frac{e^{ikr}}{\sqrt{4\pi\ell}\,r}
               \chi\big((r-L)/\ell\big)
 \label{eq:shellpacket}
\end{equation}
has norm one and kinetic residual \(O(k/\ell+\ell^{-2})\) at \(k^2/2\). Let \(J_0\) insert the normalized transverse vacuum with the physical orbit density. For a smooth compactly supported approximation to the internal ground state, the coefficients coupling this packet to transverse excitations have sizes \(O(L^{-1/2}),O(L^{-1}),O(L^{-3})\) at differential orders zero, one and two, by \eqref{ch:native-coefficients}. All slow derivatives through order two are bounded along the shells for this fixed approximation, and the smooth vacuum coefficients converge to those of the free center and the dimer. The limiting residual norm on the original matrices is therefore bounded by the internal approximation error. Taking successive rotation- and gauge-invariant graph-core approximations and choosing \(L\) large for each gives a singular Weyl sequence at \(e_b+k^2/2\), including \(k=0\). The radial motion, filled vacuum and dimer ground state together have spatial spin zero.

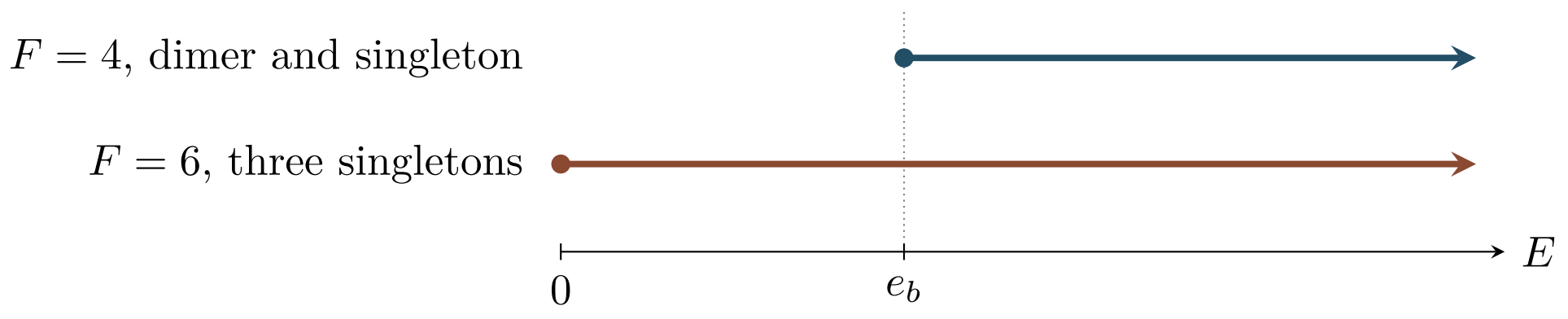
\begin{figure}[t]
\centering
\begin{tikzpicture}[x=0.93cm,y=0.75cm,>=stealth,font=\small]
\draw[->] (0,0)--(8.25,0) node[right] {\(E\)};
\draw (0,0.09)--(0,-.09) node[below] {\(0\)};
\draw (3,0.09)--(3,-.09) node[below] {\(e_b\)};
\draw[densely dotted,gray] (3,.15)--(3,2.6);
\node[anchor=east,align=right] at (-.18,2.1) {\(F=4\), dimer and singleton};
\node[anchor=east,align=right] at (-.18,.95) {\(F=6\), three singletons};
\draw[ink,line width=1.5pt,->] (3,2.1)--(8,2.1);
\fill[ink] (3,2.1) circle (2.2pt);
\draw[accent,line width=1.5pt,->] (0,.95)--(8,.95);
\fill[accent] (0,.95) circle (2.2pt);
\end{tikzpicture}
\caption{Onsets for two specified scalar escape channels in four-supercharge \(SU(3)\). At four fermions the pair--singleton channel starts at the bosonic dimer rest energy. At six fermions the three-singleton channel extends down to zero, and bound-pair channels can coexist with it.}
\label{fig:threshold}
\end{figure}

For the opposite bound the dimer must be allowed to grow with the shell radius \(R\). Away from the commuting configurations the quartic potential grows as \(R^4\) against a fermion mass of order \(R\). Near a regular commuting triple the three center gaps on a buffered angular patch are comparable to \(R\), so the two missing occupations cost at least \(cR\). It remains to cover the pair--singleton configurations. Write \(q_A\) for the closed quadratic form of a semibounded self-adjoint operator \(A\), so \(q_A[u]=\langle u,Au\rangle\) on its operator domain. On each pair--singleton patch decompose the physical wave as \(J_0p+q\), with \(q\) perpendicular to the transverse vacuum. The vacuum Hamiltonian is compared to \(h=-\Delta_z/2+H_b\) by \eqref{ch:compressed-form}, and the shifted complementary form controls \(R\|q\|^2\) as well as the slow gradient of \(q\). Since \(H_b\) is bosonic, \(q_{h+C}[p]\) controls \(\|\nabla p\|^2\) without any loss in \(R\). Absorbing the mixed form \eqref{ch:mixed-form} in a fraction of the complement therefore gives
\begin{equation}
 q_{H_G}[J_0p+q]\ \ge\
 (1-C R^{-3})q_h[p]-C R^{-2}\|p\|^2+(cR-C)\|q\|^2 .
 \label{eq:binaryform}
\end{equation}
For the zeroth-order excitation, squaring its size \(R^{-1/2}\) and dividing by its energy cost \(cR\) accounts for \(R^{-2}\|p\|^2\). The first slow derivative contributes \(R^{-3}q_{h+C}[p]\) by the same argument. A partition of unity on the finite angular cover costs \(O(R^{-2})\) in the kinetic form. The exterior variational bound thus excludes essential spectrum below \(e_b\), and the outgoing shells give
\begin{equation}
 \sigma_{\rm ess}\big(H_{SU(3)}\!\mid_{F=4,J=0}\big)
 =\sigma_{\rm ess}\big(H_{SU(3)}\!\mid_{F=4}\big)=[e_b,\infty).
 \label{eq:edge}
\end{equation}
For \(F<4\) the pair--singleton vacuum is also unavailable. The energy lower bound then diverges on every exterior patch, so these sectors have compact resolvent. At \(F=6\) the three-singleton vacuum becomes available and the scalar fully separated channel starts at zero, as illustrated in Fig.~\ref{fig:threshold}.

Figure~\ref{fig:threshold} suggests a way to see the distinction in a finite basis. A cutoff diagnostic in which \(N_B E_k\) approaches a constant is sensitive to states approaching zero as the boson cutoff \(N_B\) grows \cite{Ambrozinski}. A discretized scattering level approaching \(e_b>0\) fails that test even at arbitrarily large cutoff. One can instead project onto the internal dimer ground state and measure the spreading of the relative wave just above \(e_b\).

\section{The unbroken gauge algebra as a particle label}\label{sec:channels}

\subsection{From a commuting configuration to a channel space}
When a dimer escapes, its center separates from the singleton while its two internal colors continue to fluctuate. The corresponding distinction for any gauge group is visible by writing \(X=R\omega\), with \(|\omega|=1\). Since the bosonic potential is \(R^4 V(\omega)\), a finite-energy escape direction must approach the commuting set \(V(\omega)=0\), where all entries of a limiting tuple can be conjugated into one Cartan algebra \(\mathfrak t\). A root \(\alpha\) for which some \(\alpha(\omega_\mu)\ne0\) produces transverse frequencies of order \(R\). The roots that annihilate the tuple retain their interactions in the internal Hamiltonian.

Let \(Z\subset\mathfrak t\) be the intersection of the kernels of the roots vanishing on the tuple. We call \(Z\) a flat, include \(Z=\mathfrak t\) when no root vanishes, and retain \(Z\ne0\) for escape. A generic tuple in \(Z\otimes\R^d\) has pointwise centralizer \(L=Z_G(Z)\) with Lie algebra
\begin{equation}
                  \mathfrak l=Z\oplus\bigoplus_a\mathfrak s_a .
 \label{eq:levi}
\end{equation}
The closure of \(\exp Z\) is a torus with centralizer \(L\). In a compact connected group the centralizer of a torus is connected~\cite[Corollary~4.51]{Knapp}, so the Lie algebra in \eqref{eq:levi} accounts for the entire pointwise stabilizer. Its simple ideals \(\mathfrak s_a\) carry the internal Hamiltonians. Let \(\Hil_{\pp,f_a}(H_a)\) denote the closed span of the normalizable eigenstates of the corresponding relative physical Hamiltonian of occupation \(f_a\). The definition includes a genuine eigenstate whose energy happens to lie in another continuum. A subsystem in its own continuous spectrum can divide further and is accounted for at its final separation.

Choosing a Cartan and a flat introduces gauge copies. If \(W\) is the Weyl group of \(G\), \(W_L\) the Weyl group of \(L\), and \(N_W(Z)\) the setwise stabilizer of \(Z\), the remaining group on that flat is \(K_Z=N_W(Z)/W_L\). Its action includes the motion of the centers, the induced action on internal eigenstates, and the graded interchange of identical fermionic factors. In the following formula the action is the one obtained by transporting the filled transverse vacuum. The space associated with a flat is
\begin{align}
 \Hil_Z(F)
 &=\bigg[\bigoplus_{\{f_a\},j}^{\ \prime}
 L^2(Z\otimes\R^d)\otimes
 \Lambda^j(Z_\C\otimes\C^\nu)\otimes
       \bigotimes_a\Hil_{\pp,f_a}(H_a)
 \bigg]^{K_Z},\label{eq:channels}\\
 F&=\tfrac{\nu}{2}\big(\dim\mathfrak g-\dim\mathfrak l\big)
                         +j+\sum_a f_a,
 \qquad 0\le j\le \nu\dim Z . \label{eq:charge}
\end{align}
The prime selects the assignments at fixed \(F\). Taking the invariant part after the sum matters when a gauge transformation interchanges two fragments with different occupations. In \eqref{eq:charge} each pair of opposite stretched roots removes two real color generators from the centralizer and places \(\nu\) fermions in the filled line. The remaining center modes and the internal fermions are free to carry the indicated degrees. In \(SU(N)\), grouping equal center eigenvalues with multiplicities \(n_1,\ldots,n_k\) gives \(\nu\sum_{a<b}n_an_b\) stretched occupations. For a general group the embedding in \eqref{eq:levi} matters because isomorphic algebras can sit on inequivalent flats.

We form \(\Hil_{\rm ch}(F)=\bigoplus_{[Z]}\Hil_Z(F)\) over the nonzero Weyl orbits of flats. On each summand the comparison Hamiltonian \(h_{\rm ch}\) acts by \(-\Delta_z/2+\sum_a H_a\). The identification \(J\) inserts the normalized transverse vacuum, restores the gauge-orbit measure, and localizes initially to separated center velocities. These velocity cells, with all choices of positive margins admitted, are dense for free center motion.

\begin{theorem}[Complete particle channels at fixed group]\label{thm:channels}
For the relative theory \eqref{eq:H} at each fixed compact, simply connected simple \(G\), the ordinary incoming and outgoing wave operators for the channel space \eqref{eq:channels} summed over inequivalent flats obey
\begin{equation}
 W_\pm=\operatorname*{s-lim}_{t\to\pm\infty}
          e^{itH_G}J e^{-it h_{\rm ch}},\qquad
 W_\pm^*W_\pm=1,\qquad W_\pm W_\pm^*=P_{\ac}(H_G),
 \quad P_{\scs}(H_G)=0 .
 \label{eq:W}
\end{equation}
The identities hold at \(d=3,5\) in every conserved occupation. The two spectral projections in \eqref{eq:W} refer to the absolutely continuous and singular continuous parts of the physical Hamiltonian, respectively.
\end{theorem}

Consider a continuous wave orthogonal to every channel, with energy in a compact band away from subsystem thresholds and eigenvalues. By the propagation estimates in Appendix~\ref{app:channels}, an annulus with radii proportional to time contains all but arbitrarily small norm. There the transverse gap suppresses excitations. At intermediate center momenta we compare the remaining wave with free center motion and the interacting internal Hamiltonian. If an internal subsystem also escapes, its lower-rank channel decomposition resolves the later breakup. Differentiating the comparison across the region where both descriptions apply requires the coordinate and phase estimates in Appendix~\ref{app:overlap-jets}.

The intermediate-momentum contribution then vanishes by orthogonality. Slow centers cannot reach the annulus in logarithmic time average, while the kinetic-energy bound makes the high-momentum tail arbitrarily small. A wave of almost full norm in the annulus would thus have arbitrarily small mean weight there. This contradiction proves the range assertion on the regular bands. The exceptional set is closed and countable, and consequently carries no continuous spectral measure.

\subsection{When the same fragment separates in two steps}
Suppose a fragment recoils and then separates internally. Inserting the occupied fermions in these two stages must give the same asymptotic wave as inserting all final fragments directly. At four supercharges a canonical paired volume fixes the two phases to agree. At eight supercharges a noncommuting internal configuration can give a finite geometric phase, whose size can be computed on a pair of spin doublets.

In the eight-supercharge \(\Spin(7)\) theory an outer Cartan displacement \(R(1,1,1)\) in the fifth spatial direction leaves an \(A_2\) subsystem. A second displacement \(\eta(1,1,-2)\) in the fourth direction leaves the long-root \(A_1\) on \(e_1-e_2\). The two doublets of stretched roots have opposite shifts but unequal gaps. The doublet with gap \(R\) sees a shift \(+\eta\), and the one with gap \(2R\) sees \(-\eta\).

In the remaining \(A_1\) choose internal matrices \(uJ_i\), \(i=1,2,3\), with \(J_i=\sigma_i/2\) on a doublet. On either doublet the eigenvalues of \(\sum_i\sigma_i\otimes J_i\) are \(1/2\) and \(-3/2\) with multiplicities \(3\) and \(1\). For each eigenvalue \(\lambda\), the negative spin line belongs to a two-level system whose mass vector has components \(\lambda u\), the signed shift, and the gap \(m\). Its Berry phase around the triangle connecting the two resolutions is half the solid angle subtended at the origin, and the reversed color contributes the same phase.

For a doublet with gap \(m\) and signed shift \(\eta\), traversing the direct path first gives the phase
\begin{equation}
 \Phi(m,\eta,u)=
 \sum_{\lambda}2n_\lambda\arctan
 \frac{\lambda u\eta}
 {(\sqrt{m^2+\eta^2}+m)
  (\sqrt{m^2+\eta^2+\lambda^2u^2}+\sqrt{m^2+\eta^2})}.
 \label{eq:solid}
\end{equation}
The three overlaps in \eqref{eq:solid} follow the negative spin line through \((0,0,m)\), \((0,\eta,m)\), and \((\lambda u,\eta,m)\) and back to its starting point. The argument of their product gives the phase of one occupied spin line, and \(n_\lambda\) counts the copies at eigenvalue \(\lambda\). Within either doublet the term linear in \(u\) cancels since \(\sum_\lambda n_\lambda\lambda=0\). The cubic term is proportional to \(-\sum_\lambda n_\lambda\lambda^3=3\). Its \(m^{-4}\) dependence leaves a fraction \(1-2^{-4}\) after the oppositely shifted doublet is included, giving
\begin{equation}
 \Phi(R,\eta,u)+\Phi(2R,-\eta,u)
 =\frac{45}{128}\frac{\eta u^3}{R^4}
 + O\!\left(\frac{|\eta u^3|(\eta^2+u^2)}{R^6}\right).
 \label{eq:berryexample}
\end{equation}
The expansion holds for \(|\eta|,|u|\ll R\), and reversing the triangle reverses its sign. A pointwise small-argument expansion would be insufficient for a fragment whose own size varies with the outer radius.

For a fragment of variable size the physical estimate takes a different form. Let \(p\) be its internal and center wavefunction, \(h\) the comparison Hamiltonian, and \(m_{\rm rel}\) the phase relating the two insertions. On a shell of radius \(R\), Appendix~\ref{app:channels} gives
\begin{equation}
 \|(m_{\rm rel}-1)p\|\le C R^{-3/2}\|(h+C)^{1/2}p\|.
 \label{eq:berryest}
\end{equation}
A large nearly commuting subsystem can spread at low energy, yet the Lie bracket that generates the phase is controlled by the internal potential. The same energy bound therefore limits the phase change of the physical wave across the comparison triangle. At positive separation speed the error in \eqref{eq:berryest} is integrable in time, so the finite-radius phase of \eqref{eq:berryexample} is compatible with a unique asymptotic particle.

\section{A collision that changes the embedding}\label{sec:triality}

\subsection{Two spinor embeddings of the same fragment}\label{sec:spinor-valleys}

Consider a bosonic \(SU(n)\) ground state moving away from a collision in the four-supercharge \(\Spin(2n)\) theory, with \(n\ge4\) even. Two inequivalent embeddings give this fragment the same internal Hamiltonian and the same dispersion relation. In an oriented Cartan frame, their center lines are generated by
\begin{equation}
 w_+=\tfrac12(1,\ldots,1,1),\qquad
 w_-=\tfrac12(1,\ldots,1,-1).
 \label{eq:walls}
\end{equation}
We call the asymptotic region around the gauge and spatial orbit of either line a \emph{spinor valley}, after the two end nodes of \(D_n\). Weyl transformations permute coordinates and change an even number of signs. Reversing a line also preserves sign parity for even \(n\), so the two lines in \eqref{eq:walls} remain inequivalent under all gauge identifications. An orientation-reversing transformation of the vector representation exchanges the two valleys and identifies their internal Hamiltonians. For odd \(n\), line reversal already identifies the two choices.

For each \(a<b\), exactly one of \(e_a-e_b,e_a+e_b\) vanishes on a spinor line. The unbroken algebra is \(\mathfrak u(n)\), and the \(q=\binom n2\) stretched pairs of opposite roots require \(2q\) fermions in their transverse ground state. At occupation \(F_n=n(n-1)\) this uses all the fermions, leaving the center empty and the \(\mathfrak{su}(n)\) subsystem in degree zero. The bosonic Hamiltonian confines by \eqref{eq:bosbound}, with a unique positive ground state invariant under spatial rotations and every isometric color automorphism. Let \(e_0<e_1\) be its lowest two distinct energies on gauge singlets, in the metric induced from \(\mathfrak{so}(2n)\). The definition of \(e_1\) includes all internal spins because an excited state can combine its spin with relative orbital angular momentum to give \(J=0\).

Gauge transformations preserving a valley act on the occupied fermions through real color isometries commuting with \(\Theta=(i\sigma_2\otimes 1)\mathsf K\), where \(\mathsf K\) is complex conjugation in a real color frame. The paired volume in \eqref{ch:vacuum-phase} is equivariant under these isometries and spatial spin rotations. With the positive internal ground state, it therefore supports precisely one spin-zero orbital mode, the radial \(s\)-wave, which is even under reversal of the center line. Each of the two valleys has this single propagating mode throughout \(e_0<E<e_1\).

An asymptotic observer can distinguish the embeddings in the oriented \(2n\)-dimensional vector representation. For \(u\in S^2\) write \(X(u)=\sum_{\mu=1}^3u_\mu X_\mu\) as a real skew matrix, and use normalized spherical average \(\langle\cdot\rangle_u\) to define
\begin{equation}
 {\cal N}_n(X)=\left\langle
        \big(-\tfrac12\Tr X(u)^2\big)^{n/2}\right\rangle_u,\qquad
 {\cal P}_n(X)=\frac{\langle\Pf X(u)\rangle_u}{{\cal N}_n(X)} .
 \label{eq:tags}
\end{equation}
For \(X\ne0\) at least one spatial combination is nonzero on an open set of directions, so the denominator is positive. A radial cutoff extends the ratio smoothly across the origin. On a ray in a center line \(w\), all skew blocks are proportional to \(w_a\) with the same spatial factor. This factor cancels from the ratio, giving \({\cal P}_n=\prod_a w_a/|w|^n\). Along the outgoing ground mode, the internal state stays localized and the normal oscillators have width \(O(R^{-1/2})\) at center separation \(R\). Consequently the reading converges in probability to \(+n^{-n/2}\) or \(-n^{-n/2}\). Reversing the orientation convention interchanges the labels.

The two readings are extremal among commuting configurations. If \(v_a\in\R^3\) denotes the three entries of the \(a\)-th Cartan block, the arithmetic--geometric mean inequality gives
\(\big|\prod_a(u\cdot v_a)\big|\le n^{-n/2}
(\sum_a(u\cdot v_a)^2)^{n/2}\).
Equality after averaging requires all \((u\cdot v_a)^2\) to agree for every \(u\). The rank-one forms \(v_av_a^t\) are then equal, so all \(v_a=\pm v_1\). Their sign parity distinguishes the two spinor valleys. At \(F_n\) no other admissible flat contains either spinor line. By \eqref{eq:charge} an admissible flat must retain at least \(q\) positive roots. If it contains a spinor line, its retained roots are among the \(q\) that vanish on the line. It must therefore retain all \(q\), whose common kernel is the line itself. Since the root arrangement is finite, the remaining admissible flats on the unit sphere stay a positive distance from both extremal orbits. Sufficiently small windows around the two Pfaffian readings therefore isolate the corresponding ground channels in this band.

\subsection{Transmission between two selected valleys}\label{sec:pair-transmission}

\begin{theorem}[Conversion of the spinor embeddings]\label{thm:spinor}
Fix even \(n\ge4\) and restrict the four-supercharge \(\Spin(2n)\) theory to \(F=F_n\), \(J=0\). For either ordered pair of distinct spinor valleys \(i,j\), the energy-normalized ground-mode amplitude satisfies
\begin{equation}
                 S_{ji}^{(F_n)}(E)\ne0
           \quad\text{for almost every }e_0<E<e_1 .
 \label{eq:spinor-conversion}
\end{equation}
Thus a unit packet \(f\) incident in valley \(i\), supported in this band, has outgoing probability \(\int |f(E)|^2|S_{ji}^{(F_n)}(E)|^2\,dE>0\) in valley \(j\).
\end{theorem}

In \(\Spin(12)\), an incident bosonic \(SU(6)\) fragment at \(F=30\) can change its Pfaffian reading from \(+1/216\) to \(-1/216\). The reflection exchanging the embeddings also commutes with the identity scattering matrix, so symmetry allows perfect return. Even mixing at finite distance can give this outcome. For example, conjugate a free Hamiltonian on several radial channels by a position-dependent unitary that mixes them on a bounded interval and equals one outside. Freely moving packets leave the interval in either time direction, so both wave operators equal the same unitary and the scattering matrix is the identity. A signal at finite distance alone cannot establish conversion.

For energies in a compact interval \(I\subset(e_0,e_1)\), we isolate the two valleys in disjoint cones around their orbits on the unit sphere of \(\mathfrak{so}(2n)^3\). We impose Dirichlet conditions on the sides and on the caps at \(\rho=R_0\), where \(\rho=|X|\) and the common radius \(R_0=R_0(I,n)\) is chosen sufficiently large. Within each cone the nonzero restricted roots have only two opposite charges, which excludes a neutral cubic in the normal coordinates. The transverse gap grows with \(\rho\), and an internal displacement of order \(\rho\) costs a diverging energy by \eqref{eq:bosbound}. Appendix~\ref{app:transmission} uses these bounds to construct the angular ground mode on the sphere of \emph{physical} radius \(\rho\) and to put the angular energy of every perpendicular state above \(I\).

With this gap, the solution \(\phi_i(z)\) with unit radial ground-mode slope at the cap and controlled exponential growth continues from \(z=-s<0\) through the threshold. Its ground component at \(E=e_0+k^2/2\in I\) has asymptotics
\[
 a_i(\rho,E)=A_{i,+}(E)e^{ik\rho}+A_{i,-}(E)e^{-ik\rho}+o(1).
\]
Reality gives \(A_{i,-}=\overline{A_{i,+}}\), and comparison with \(\sin(k(\rho-R_0))/k\) keeps both coefficients away from zero on \(I\) for large \(R_0\).

At negative energy, the fermion transport along each Cartesian heat path preserves quaternionic pairs. The matrix element between paired volumes is a product of squared singular values of the compressed one-body transport, while the allowed color motions give full-rank overlap on a set of paths of positive measure. This sign enters the asymptotic experiment through the boundary current of \(\phi_i(-s)\). On the individual cone, pairing a compact source with \(\phi_i(-s)\) measures the unique obstruction to faster decay of its Dirichlet response. Positivity of that response fixes the sign of the current, and the unit slope gives a nonzero current on a patch of each cap.

Choose \(\chi_i\) vanishing near the Dirichlet faces and equal to one far along the interior. For \(H=H_{\Spin(2n)}\) in the selected sector put \(d_i(z)=(H-z)(\chi_i\phi_i(z))\). The full resolvent now enters the scalar function
\begin{equation}
 {\cal F}_{ji}(z)=\big(d_j(z),(H-z)^{-1}d_i(z)\big)_{\rm bil},
 \label{eq:analytic}
\end{equation}
where the pairing is the complex bilinear extension from the real fermion space. At \(-s\) the disjoint supports allow us to replace both smooth defects by their boundary currents. The strictly positive heat pairing of the two cap patches yields \({\cal F}_{ji}(-s)>0\) with consistently oriented ground modes.

At positive energy, launch a packet from valley \(i\) and pair its retarded evolution with an outgoing packet in valley \(j\). The two defects have disjoint supports, so their direct overlap vanishes. The remaining term propagates through the full Hamiltonian and has the upper boundary value
\begin{equation}
 {\cal F}_{ji}(E+i0)
 =ik\,\overline{A_{j,+}(E)}\, S^{(F_n)}_{ji}(E)\, A_{i,-}(E)
 =ik\,A_{j,-}(E)\, S^{(F_n)}_{ji}(E)\, A_{i,-}(E).
 \label{eq:boundary}
\end{equation}
The outgoing coefficient is conjugated because the inner product is antilinear in the outgoing packet. Both channels use the free radial normalization \(\sqrt{2/(\pi k)}\sin(kr)\) per unit energy. Although the packets select two valleys, the resolvent in \eqref{eq:analytic} includes propagation through every open channel.

Equations~\eqref{eq:analytic} and~\eqref{eq:boundary} give a pairwise criterion for conversion. The isolated ground mode and the gap to perpendicular waves in each cone select nonzero traveling coefficients, while the signed boundary currents give a nonzero value below the spectrum. Appendix~\ref{app:transmission} identifies the upper boundary value in \eqref{eq:boundary} as a nontangential limit almost everywhere. If \(S_{ji}^{(F_n)}\) vanished on a set of positive measure, boundary uniqueness~\cite{Privalov} would force \({\cal F}_{ji}\) to vanish throughout its connected domain, contradicting its value at \(-s\). Applying the argument on compact intervals exhausting the band proves Theorem~\ref{thm:spinor}.

\subsection{The three triality probabilities}\label{sec:triality-prob}

At \(n=4\), a third embedding allows us to determine how the amplitudes of this reaction are related. The twelve opposite-root pairs of \(D_4\) would use twenty-four fermions at full separation. At \(F=12\) at least six pairs must remain unstretched. Removing an outer node of the Dynkin diagram retains \(A_3\), with six positive roots, while removing its middle node retains \(A_1^3\), with only three. Every smaller Levi subsystem retains fewer than six. The only possible separating lines are therefore \(w_+\), \(w_-\), and a vector line \(w_{\rm v}=(1,0,0,0)\). Triality permutes the three valleys and gives the same bosonic \(SU(4)\) ground state on each.

A second quartic reading measures how the Cartan block norms are distributed,
\[
 {\cal T}(X)=\frac{\langle\tfrac12\Tr X(u)^4\rangle_u}{{\cal N}_4(X)}.
\]
On a center line it gives \(\sum_a w_a^4/|w|^4\). Table~\ref{tab:tags} gives the two invariant readings in each outgoing channel.
\begin{table}[htbp]\centering
\begin{tabular}{lcc}
\toprule
Embedding of the bosonic \(SU(4)\) & \({\cal T}\) & \({\cal P}_4\)\\
\midrule
Vector \(w_{\rm v}\) & \(1\) & \(0\)\\
Spinor \(w_+\) & \(1/4\) & \(1/16\)\\
Conjugate spinor \(w_-\) & \(1/4\) & \(-1/16\)\\
\bottomrule
\end{tabular}
\caption{Limiting readings of the quartic observables in the three triality valleys, for a fixed orientation of the vector representation.}
\label{tab:tags}\end{table}

Let \(\Pi\) and \(\Pi_{\rm ch}\) select \(F=12,J=0\) in the physical and channel spaces. A rotation or diagram automorphism carries each physical ground-state insertion into the insertion for the transformed channel, with the paired volume fixing its phase. The wave operators therefore intertwine these symmetries. On \(I\) the comparison space selected by \(\Pi_{\rm ch}\) is \(L^2(I,dE;\C^3)\), and the two wave operators of Theorem~\ref{thm:channels} map it unitarily onto the same physical spectral subspace \({\bf1}_I(H)\Pi\Hil_{\ac}(H)\). Triality acts in each energy fiber by the ordinary permutation representation. With \(\boldsymbol{1}=(1,1,1)^t\), the orthogonal projectors onto its singlet and doublet are \(P_1=\boldsymbol{1}\boldsymbol{1}^{\,t}/3\) and \(P_2=1-P_1\). Schur's lemma and unitarity give
\begin{equation}
 S(E)=a(E)P_1+b(E)P_2,\qquad |a|=|b|=1,\qquad
 S_{ii}=(a+2b)/3,\quad S_{ji}=(a-b)/3\quad(i\ne j).
 \label{eq:schur}
\end{equation}

\begin{corollary}[Conversion and return under triality]\label{thm:triality}
For a normalized packet \(f\) incident in one of the three valleys, supported in the first band, put \(\delta(E)=\arg(a(E)/b(E))\). Its probabilities are
\begin{equation}
 \begin{aligned}
 \Pr(i\to j)&=\int_{e_0}^{e_1}|f(E)|^2\,
       \frac{2-2\cos\delta(E)}{9}\,dE>0\qquad (j\ne i),\\
 \Pr(i\to i)&=\int_{e_0}^{e_1}|f(E)|^2\,
       \frac{5+4\cos\delta(E)}{9}\,dE\ge\tfrac19 .
 \end{aligned}
 \label{eq:packetprob}
\end{equation}
\end{corollary}
An incident vector fragment has weights \(1/3\) and \(2/3\) in the singlet and doublet. The relative phase accumulated by these components gives the equal probabilities for the two spinor readings in \eqref{eq:packetprob}. By Theorem~\ref{thm:spinor} the two eigenphases differ almost everywhere. Even for a relative phase of \(\pi\), at least one ninth of the flux returns to the incident species. This last bound uses the complete three-dimensional fiber at \(n=4\).

\subsection{Resolving the internal and relative fermions}\label{sec:descendants}

A different preparation reveals which degrees of freedom carry two additional fermions after the embedding changes. Return to either spinor valley at general even \(n\), and let \(v_i\) denote its normalized ground mode at \(E\in(e_0,e_1)\). On the full gauge-singlet space the physical charges obey \(\{\mathcal Q_\alpha,\mathcal Q_\beta^\dagger\}=2\delta_{\alpha\beta}H\), with \(\mathcal Q_\alpha^\dagger\) raising the fermion number \(F\) by one. Dividing the raising charges by \(\sqrt{2E}\) gives dimensionless operators on this positive-energy multiplet. On a separated fragment they act as
\[
 a_\alpha^\dagger=\sqrt{x}\,b_\alpha^\dagger+\sqrt{y}\,c_\alpha^\dagger,
 \qquad x=\frac{e_0}{E},\qquad y=\frac{E-e_0}{E},\qquad \alpha=1,2 .
\]
The internal and relative charges \(b^\dagger,c^\dagger\) are normalized at their respective energies, with \(c^\dagger\) containing the free spin matrix \(\sigma\cdot\widehat p\). The lowering charges annihilate \(v_i\), and the two sets of raising operators anticommute on the graded product. The scalar double raise therefore gives a normalized state at \(F_n+2\),
\begin{equation}
 D_i=a_1^\dagger a_2^\dagger v_i
     =x B_i+\sqrt{2xy}\,M_i+y C_i .
 \label{eq:descendant}
\end{equation}
Here \(B_i\) has two extra fermions on the fragment, \(C_i\) has both in the relative motion, and \(M_i\) is the normalized mixed component with one in each. The three are orthogonal by internal fermion number. The mixed component includes one unit of relative orbital angular momentum. A gauge transformation reversing the center reverses both its momentum and its fermions, so each of these components respects the residual gauge identification.

To turn the decomposition \eqref{eq:descendant} into outgoing probabilities, we show that both wave operators intertwine the physical and asymptotic supercharges. Let \(\mathcal Q\) denote a raising or lowering charge, \(\mathcal Q_h\) its separated counterpart, and \(J\) the ground-state insertion. On a shell of radius \(R\), the leading transverse charge is \(O(\sqrt R)\) on oscillator moments and annihilates the vacuum at internal background \(U=0\). Moving to \(U\) changes the vacuum by \(O(|U|/R)\) in those norms. The commutator \([U,W]\) and the longitudinal derivative eliminated by the gauge constraint then contribute \(O(\langle U\rangle/\sqrt R)\), since \(W=O(R^{-1/2})\) on these states. The quadratic term \([W,W]\) and derivatives of the frame and orbit density contribute \(O(R^{-1})\). The retained commutator \([U,U]\) is already in the internal charge. The resulting bound on \(\mathcal QJ-J\mathcal Q_h\) is \(C R^{-1/2}\) times finitely many weighted norms of the packet and its first derivatives. Those norms are finite for the ground state and its superpartners by confinement and the polynomial first-order form of the charges.

For a smooth packet with momentum in a compact subset of \(k>0\), free propagation makes the error tend to zero in both time directions. Off the separating rays the packet decays by nonstationary phase. To pass the charge through the wave-operator limit, take the symmetric combination \(R_\alpha=\mathcal Q_\alpha+\mathcal Q_\alpha^\dagger\). It preserves the smooth compactly supported singlet core of Section~\ref{sec:dimer} and satisfies \(R_\alpha^2=2H\) there by nilpotence and the anticommutator. Since \((R_\alpha-i)(R_\alpha+i)=2H+1\) in either order on the core, both \(R_\alpha\pm i\) have dense range. Its self-adjoint closure therefore squares to \(2H\), has domain \(D(H^{1/2})\), and commutes with \(e^{itH}\).

The packet \(e^{itH}Je^{-ith_{\rm ch}}f\) tends to \(W_\pm f\), and the estimates for the two charges give convergence of its image under \(R_\alpha\) to \(W_\pm R_{\alpha,h}f\), where \(R_{\alpha,h}\) is the separated symmetric charge. Closedness gives \(R_\alpha W_\pm f=W_\pm R_{\alpha,h}f\). For a packet of degree \(F\), projection onto degree \(F+1\) or \(F-1\) then gives the required intertwining for the raising or lowering charge, since \(W_\pm\) preserves fermion number. Energy intertwining on compact subintervals allows the same passage for the normalized charges and their products on the finite multiplet.

Now prepare \(D_i\) and resolve the outgoing \(F_n+2\) state on valley \(j\) into \(C_j,M_j,B_j\). Moving the two raises to the outgoing bra lowers these three states to \(y v_j,\sqrt{2xy}\,v_j,xv_j\), respectively, in the phases of \eqref{eq:descendant}. Their probabilities at energy \(E\) are therefore
\begin{equation}
 \big(\Pr(D_i\to C_j),\Pr(D_i\to M_j),\Pr(D_i\to B_j)\big)
   =|S^{(F_n)}_{ji}(E)|^2\big(y^2,\,2xy,\,x^2\big).
 \label{eq:resolved}
\end{equation}
All three are positive for almost every energy in the band. The same calculation with \(C_i\) incident gives \(y^2|S^{(F_n)}_{ji}|^2\) into the specified outgoing state \(D_j\), and hence at least this probability into valley \(j\) inclusively. The coherent input \((xC_i-yB_i)/\sqrt{x^2+y^2}\) has zero amplitude into \(D_j\) because its double lowering vanishes. Other outgoing polarizations can still carry flux from this input. For packets, the coefficients in the preparation are evaluated at each energy, and the joint probabilities in \eqref{eq:resolved} are averaged against the incident energy density.
\section{Discussion}\label{sec:discussion}

An \(SU(n)\) fragment can emerge from a collision in the same bosonic ground state and with a different embedding in the color algebra. In the four-supercharge \(\Spin(2n)\) theory at even \(n\ge4\), the two spinor embeddings give opposite extremal Pfaffian readings at large separation. A switch from one extreme to the other records conversion in either orientation convention, since reversing the color orientation changes both signs.

For a doubly raised incoming state, measuring the internal fermion number of the outgoing fragment also reveals how the two extra fermions divide between the fragment and its relative motion. The three probabilities in \eqref{eq:resolved} share the factor \(|S_{ji}^{(F_n)}(E)|^2\), so their relative weights can be measured at fixed energy without knowing the absolute conversion probability. Among these three outcomes in valley \(j\), the fraction with both extra fermions on the fragment is \((e_0/E)^2\) and tends to one as \(E\downarrow e_0\). A bound on the absolute conversion probability at threshold would require estimates uniform as the relative momentum tends to zero.

In four-fermion \(SU(3)\), all four fermions occupy the modes connecting the escaping dimer to the singleton. The dimer is therefore bosonic, and its positive ground energy \(e_b\) sets the continuum threshold. In a finite-basis calculation, projecting onto this internal ground state and tracking the spread of the relative wave as the basis grows would help distinguish the discretized dimer continuum from a state localized in all three colors.

The eight-supercharge \(\Spin(8)\) model offers a direct comparison. At \(F=24\) the three triality-related center lines again support a bosonic \(SU(4)\) ground fragment. The four-supercharge sign argument uses an antiunitary pairing preserved by the one-body mass along every matrix history. In five spatial dimensions the spin conjugation preserves each Clifford matrix, while conjugating the color factor sends \(i\ad X_\mu\) to \(-i\ad X_\mu\). The resulting antiunitary map exchanges the positive and negative mass spaces. It therefore supplies no orientation of the occupied space of the kind used in Appendix~\ref{app:transmission}. The finite Berry phase of the eight-supercharge \(\Spin(7)\) example in \eqref{eq:berryexample} illustrates the difference. Whether the eight-supercharge collision has nonzero triality conversion remains open.

Even with four supercharges the size of the conversion amplitude is unknown. In \(\Spin(8)\) it can be obtained from the difference of the singlet and doublet scattering phases. Imposing equivalent distant radial boundaries in the three valleys makes the free propagation phase common to both quantization conditions. Finite-volume phase extraction has been demonstrated for two colors in the sixteen-supercharge model~\cite{BajajLin}. Controlling the difference of the finite-radius errors here as \(k\downarrow0\) would decide whether the first embedding-changing signal is already present in the scattering lengths.

\section*{Acknowledgments}
\addcontentsline{toc}{section}{Acknowledgments}
GPT-6.2 Astra contributed arguments for geometric phase control through successive breakups and for proving nonzero transmission between spinor valleys by analytic continuation from negative energy; GPT-6.2 Astra and Gemini-4 Neon were used to assist with writing and editing the manuscript.

\appendix
\section{Channel comparison through successive breakups}
\label{app:channels}

A fragment can separate from its neighbors while its own matrices are still
interacting. We keep its full internal Hamiltonian in the comparison
dynamics until its bound and scattering parts can be separated by induction.
If it breaks up, the same final configuration admits two sets of transverse
coordinates. The slow kinetic energy differentiates the change between
them in every internal direction, including directions transverse to the
dilation used in the propagation estimate.

We use the range argument for sixteen-supercharge \(SU(N)\) in
Ref.~\cite{Qi}, Secs.~4--5. Its local hypotheses must be checked for the
restricted roots and the occupied fermion space of the present model.
We derive the gap and the change of the occupied line below, including
the ordinary derivatives required by the second commutator. The second
change concerns propagation in three spatial dimensions, where the scalar
partial wave requires a different bounded function of the radial dilation.
The constants may depend on the group, the fixed angular apertures and
the energy cutoffs. We use \(g=1\).

\subsection{The energy of a fluctuating fragment}

Near a commuting tuple, a generic linear combination of its entries has
the same vanishing roots as the tuple itself. Its centralizer thus gives
the flat \(Z\) of Section~\ref{sec:channels}. In the complexified complement
of \(\mathfrak l=\operatorname{Lie}Z_G(Z)\), collect the root spaces with
the same nonzero character \(\chi\) of the central torus in a single
space \(V_\chi\). The retained matrices can mix all the colors in this
space while leaving the character unchanged.

Write \(z_i\in Z\) for the centers, \(U_i\in[\mathfrak l,\mathfrak l]\)
for the retained matrices and \(W_i\) for the normal coordinates. On a
strict chart at radius \(R\), all vectors
\(r_\chi=(\chi(z_i))_{i=1}^d\) have length at least \(\kappa R\), with
small fixed apertures in \(|U|<\epsilon R\) and \(|W|<\eta R\).
We remove the longitudinal gauge direction by imposing
\(r_\chi\cdot W_\chi=0\) on the pair \(\chi,-\chi\).
Let \(Y_i\) be the Hermitian infinitesimal action of \(U_i\) on \(V_\chi\)
and \(D=|r_\chi|+Y_n\), where \(n=r_\chi/|r_\chi|\).
Eliminating the longitudinal gauge variation in the kinetic energy gives
the inverse transverse mass \(G\), while the quadratic variation of the
commutator potential gives the stiffness \(K\). With transverse indices
\(I,J,A\perp n\), their entries are
\begin{align}
 G_{IJ}&=\delta_{IJ}+Y_I D^{-2}Y_J,\nonumber\\
 K_{IJ}&=\delta_{IJ}\bigl(D^2+\textstyle\sum_A Y_A^2\bigr)
                  +Y_IY_J-2Y_JY_I .                    \label{ch:frozen}
\end{align}
In the first formula the longitudinal orbit row is \(D\) and the
transverse row is \(Y_I\). In the second formula the term involving the
background bracket comes from pairing \([U_i,U_j]\) with
\([W_i,W_j]\) and using invariance of the color metric. The matrices
\(G,K/R^2\) are uniformly positive for sufficiently small \(\epsilon\).
The fermion mass is the Clifford contraction of \(r_\chi+Y\) and is
gapped at zero. Filling its negative eigenspace in both root orientations
uses \(d-1\) fermions per opposite root pair and leaves a unique
transverse ground line. The first transverse excitation above that line costs at least \(cR\).

\paragraph{Energy and phase of the transverse ground state.}
Let \(e_{\mathrm f}\) be the bosonic zero-point energy plus the filled
fermion energy, summed over the nonzero characters, and \(V_{\rm ret}\)
the potential of the retained matrices. With the inverse mass in
\eqref{ch:frozen}, the two energies cancel for mutually commuting \(Y_i\)
because after simultaneous diagonalization each weight describes an oscillator
normal to the displaced center vector. The relevant remainder for a
noncommuting background is
\begin{equation}
             |e_{\mathrm f}(z,U)|\ \leq\ C R^{-3}V_{\rm ret}(U).
             \label{ch:scalar}
\end{equation}
Expand the positive oscillator square root and the absolute fermion
mass in \(Y/|r_\chi|\) on a slightly larger tube. In the fermion expansion, \(|M|=\sqrt{M^2}\) expresses the spin
coefficients as traces of even Clifford words. Moving the first generator around such
a trace by anticommutation and cyclicity reproduces the trace with a
minus sign, together with contractions multiplying shorter even traces.
Induction thus expresses every spin coefficient as a real sum of products
of Kronecker deltas. This argument applies also to the two-component
Pauli matrices.

The cancellation on commuting backgrounds removes the sorted monomial
in each group of color words with the same letters. Each remaining
term contains a commutator \(C_{ij}=[Y_i,Y_j]\). If \(A\) is the
product of the remaining Hermitian letters, the real part of the color
trace is
\[
 \operatorname{Re}\operatorname{tr}(C_{ij} A)
      =\tfrac12\operatorname{tr}\bigl(C_{ij}(A-A^\dagger)\bigr).
\]
Reordering \(A^\dagger\) back to \(A\) expresses \(A-A^\dagger\) as
a sum of terms, each containing a commutator of two letters. This
commutator and \(C_{ij}\) both represent Lie brackets of the retained
matrices, so Cauchy--Schwarz in the finite color trace bounds their product by
\(C V_{\rm ret}\). Their four letters contribute \(|r_\chi|^{-4}\)
against the energy scale \(|r_\chi|\), giving \eqref{ch:scalar}.
On the smaller tube the convergent series absorbs the polynomial
number of swaps in each degree, also after two differentiations.

Along the ground line the slow kinetic energy also sees the phase of
the occupied fermions. For four supercharges take
\(\Theta=(i\sigma_2\otimes 1)\mathsf K\), with complex conjugation
\(\mathsf K\) in a real adjoint basis. Both factors of
\(\sigma_i\otimes i\ad X_i\) change sign under the corresponding
conjugations, so the full mass commutes with \(\Theta\) and
\(\Theta^2=-1\). If \(v_j,\Theta v_j\) form an orthonormal occupied
frame, the corresponding determinant has connection
\begin{equation}
 \sum_j\bigl(\langle v_j,dv_j\rangle+
       \langle\Theta v_j,d(\Theta v_j)\rangle\bigr)
 =2\operatorname{Re}\sum_j\langle v_j,dv_j\rangle=0 .
 \label{ch:vacuum-phase}
\end{equation}
A change of quaternionic orthonormal basis has unit complex determinant.
The paired volumes therefore define a single parallel section, multiply
under orthogonal sums, and are transported exactly by real color
automorphisms and spin rotations commuting with \(\Theta\).

For eight supercharges the spin conjugation commutes with each of the
five Clifford generators. Including color conjugation identifies the
positive space in one orientation with the negative space in the
opposite orientation. Conjugation reverses determinant curvature and
the complementary space has the opposite trace curvature, so the two
occupied contributions add. Choose the parallel section at \(U=0\)
and transport it along \(tU\), \(0\le t\le1\), at fixed \(z\).
In this gauge the connection one-form \(\mathcal A\) satisfies
\begin{equation}
 |\mathcal A_U|\leq C R^{-3}\sqrt{V_{\rm ret}},\qquad
 |\mathcal A_z|\leq C R^{-4}|U|\sqrt{V_{\rm ret}} .       \label{ch:connection}
\end{equation}
Radial transport expresses the connection as an integral of the
determinant curvature, evaluated on \(U\) and \((\delta z,t\delta U)\).
In the contour expansion of the occupied projector for the divided mass
\(M/|r_\chi|\), cyclicity puts the distinguished variation last in each
color trace. For scalar color letters the determinant curvature
vanishes, since the negative eigenspace of the five-dimensional
Clifford contraction is quaternionic. Sorting the remaining letters
therefore leaves a bracket of two \(Y\)'s. For a center variation the
distinguished letter is scalar, and the imaginary part of a trace
with only two Hermitian color letters vanishes. One further \(Y\)
is needed in that component. The factor \(R^{-1}\) for a unit
variation, the inverse power per retained letter, and convergence on
the buffered tube yield \eqref{ch:connection}.

\paragraph{Projection onto the ground state and its complement.}
Write \(h=-\Delta_z/2+\sum_a H_a\) for the comparison Hamiltonian on
the full singlet internal spaces and \(J\) for insertion of the fast
ground line with the physical orbit density. We normalize the column
in each fiber after cutting off in \(W\) and keep cutoffs in the slow
variables outside \(J\), so \(JJ^*\) is an orthogonal projection.
On the buffered shell the internal Yukawa multiplier has magnitude at most
\(CR\). Writing \(T_s\) for the kinetic energy in \(s=(U,z)\), we have
\(T_s+V_{\rm ret}\leq h+CR+C\) there and hence
\begin{equation}
 \|R^{-1/2}\nabla p\|+\|R^{-1/2}\sqrt{V_{\rm ret}}\,p\|
       \leq C\|(h+C)^{1/2}p\| .                         \label{ch:form}
\end{equation}
The factor \(R^{-1/2}\) is needed because an internal wave can spread
along a nearly commuting valley where the negative Yukawa energy grows.
For the purely bosonic fragment of Section~\ref{sec:dimer}, the internal
kinetic and potential energies are separately positive and control
both unweighted norms.

The cubic potential, linear Yukawa term in \(W\), and linear normal
metric correction have odd Gaussian parity, and a single omitted gauge
generator carries a nonzero central character. Their leading
expectations in \(Jp\) vanish. The slow metric differs from the identity
by \(O(W^2/R^2)\), whose mean is \(O(R^{-3})\). Its slow--normal cross
term yields a real first derivative whose divergence is \(O(R^{-2})\)
in the physical half-density. Writing \(i\mathcal A\) for the connection
of the unit section, the projected kinetic derivative is
\(\nabla+i\mathcal A\), and its real scalar correction is
\(O(R^{-2})\) in addition to \eqref{ch:scalar}. These metric orders
also follow directly from the slice kinetic energy in
\eqref{ch:graded-H} below. Since \(|U|\leq\epsilon R\),
\eqref{ch:connection} bounds the connection term by the product of
\(CR^{-3}\|\sqrt{V_{\rm ret}}p\|\) and \(\|\nabla p'\|\), and its
adjoint by the same expression with \(p,p'\) interchanged. Together
with \(T_s+V_{\rm ret}\leq h+CR+C\) this gives, on the buffered shell,
\begin{equation}
 \begin{aligned}
 |q_{H_G}(Jp,Jp')-q_h(p,p')|&\leq C n_R(p)n_R(p'),\\
 n_R(p)^2&=R^{-3}q_{h+C}[p]+R^{-2}\|p\|^2 .
 \end{aligned}                                           \label{ch:compressed-form}
\end{equation}
Here \(q_H(\cdot,\cdot)\) denotes the polarized energy form. In applying
\eqref{ch:compressed-form} one can take any internal form vector
supported on the shell, even if its subsystem is itself in the
continuous spectrum.

On the orthogonal complement \(Q=1-JJ^*\), an excited fast state can carry
color and leave the internal fermion energy negative. Its magnitude is at
most \(c_1\langle U\rangle\). By choosing the aperture in
\(|U|<\epsilon R\) after \(c_1\), we can absorb this cost in part of the
gap \(cR\). If \(H_Q\) denotes the compression to the complement, the
shifted form \(\Lambda_Q[q]=q_{H_Q+C}[q]\) therefore controls both
\(R\|q\|^2\) and \(\|\nabla_{\rm slow}q\|^2\).

The corresponding graph estimate also controls \(R(\mathcal N+1)q\),
\(\sqrt R\,\nabla_{\rm slow}q\) and \(V_{\rm ret}q\), where \(\mathcal N\)
counts transverse excitations. The covariant square used in
Ref.~\cite{Qi}, Sec.~4.2 and App.~B.2, applies here because the
dimensionless oscillator coefficients in \eqref{ch:frozen} are smooth
and uniformly positive. The quadratic transport of its moving frame has
coefficients \(O(R^{-1})\) and curvature \(O(R^{-2})(\mathcal N+1)\).
On the actual normal support \(|W|\le\eta R\), the cubic and quartic
remainders are bounded by \(C(\eta+\eta^2)R^2|W|^2\). Taking this bound
before applying the oscillator estimate makes their relative graph bound
small. For the slow operator the only scalar error from squaring the
kinetic plus quartic terms is \(-\frac12\int(\Delta V)|f|^2\).
Since \(|\Delta V|\le C(1+|X|^2)\) and the fermion multiplier grows
at most linearly, \(f=(1+|X|^2)^{-1/2}u\) gives the same weighted
Hessian estimate as Eqs.~(4.3)--(4.6) of Ref.~\cite{Qi}. These arguments
apply to each fixed restricted-root multiplicity.

For the mixed form, the coefficients of zero, one and two slow derivatives
in \(QH_GJ\) have orders \(R^{-1/2},R^{-1},R^{-3}\), respectively, by
\eqref{ch:native-coefficients}. In the first two terms the gap gives
\(\|q\|\le C R^{-1/2}\Lambda_Q[q]^{1/2}\). In the last term we
integrate one slow derivative by parts. If it lands on \(q\) the
complementary form bounds the gradient, while a derivative on the
coefficient gains another inverse power of \(R\). Hence, for \(R\ge1\),
\begin{equation}
 |q_{H_G}(q,Jp)|\leq C\Lambda_Q[q]^{1/2}
 \left(R^{-1}\|p\|+
       R^{-3/2}\|\nabla p\|\right).             \label{ch:mixed-form}
\end{equation}
Nested cutoffs keep the integration by parts inside the buffered chart,
so the inequalities extend from smooth compactly supported singlets by
form closure. On the graph domain one applies the two rows of the
source/complement decomposition of \(H_G\) to a localized vector \(ju\).
The slow Hessian with one inverse power of \(R\) controls the order-two
term in \(QH_GJ\), while in the adjoint row the derivatives also act on
its coefficients, with the bounds in \eqref{ch:native-coefficients}.
Absorbing these small terms at large \(R\) gives
\(\|(h+C)J^*ju\|+\|(H_Q+C)Qju\|\le C_j(\|H_Gu\|+\|u\|)\),
in the order of Eq.~(4.8) of Ref.~\cite{Qi}. The localization
commutator costs one weighted gradient on the larger chart.

\subsection{Comparing the two resolutions of a bound subsystem}

On the section \(W=0\) of a refined chart the old and new transverse
slices are generally oblique. In a refined character let \(A_i=r_i+Y_i\)
be the orbit row and \(n\) the axis defining the target slice. The linear
map between slices is \((Lw)_i=w_i-A_i(n\cdot A)^{-1}(n\cdot w)\).
The implicit gauge rotation is \(O(|w|/R)\). Its first action on the
background has nonzero central character and hence zero retained
projection. The first displacement of a retained coordinate is a
quadratic form \(\Gamma_2(w)=O(|w|^2/R)\). Taking the second derivative
of the potential along this change of coordinates gives
\begin{equation}
 K_{\rm successive}=L^*K_{\rm direct}L+E,\qquad
 (w,Ew)=2\,dV_{\rm ret}(U)[\Gamma_2(w)].                \label{ch:hessian}
\end{equation}
The derivative of the quartic potential has size \(C|U|\sqrt{V_{\rm ret}}\).
After identifying the two kinetic metrics, let \(\Gamma_{\mathrm f}\)
be the positive Gaussian precision, so
\(\Gamma_{\mathrm f}G\Gamma_{\mathrm f}=K\). Conjugating by \(G^{1/2}\) reduces its
variation to that of a positive square root with eigenvalues comparable
to \(R\). Inverting the Sylvester map \(X\mapsto SX+XS\) for this
square root \(S\) costs \(C/R\). Dividing by the unperturbed precision
costs another \(C/R\), so the relative change is bounded by
\(C|U|\sqrt{V_{\rm ret}}/R^3\). The typical displacement
\(\Gamma_2(W)\) of the retained wave has size \(O(R^{-2})\) and acts on
one slow derivative. By \eqref{ch:form}, both errors are controlled on the source form domain.

The two descriptions can also differ by parallel transport of the
filled fermions. For four supercharges the paired section in
\eqref{ch:vacuum-phase} respects direct sums, hence its relative phase
is exactly one. For eight, let \(d'\) be the shift in a character's
center vector \(r\) upon refinement. The three vertices in the
comparison are \((r,0),(r+d',0),(r+d',Y)\). The character spaces
split into their refined components throughout this triangle. Its
first leg uses the parallel section over commuting centers and its second follows
the defining radial transport in \(Y\). On the diagonal
\((r+td',tY)\) the connection annihilates the radial internal direction,
so only \(\mathcal A_z[d']\) contributes. Integrating
\eqref{ch:connection} gives the continuous choice of phase \(m\) with
\begin{equation}
 |\arg m|\leq C \frac{|d'|\,|U|\sqrt{V_{\rm ret}}}{R^4}.
 \label{ch:berry}
\end{equation}
Since \(|m-1|\leq|\arg m|\) and \(|U|+|d'|\le CR\), the estimate for
\(\|\sqrt{V_{\rm ret}}p\|\) in \eqref{ch:form} gives the
\(R^{-3/2}\) bound on the relative phase in \eqref{eq:berryest}.
Adding the Gaussian precision and displacement errors in
\eqref{ch:hessian} gives the same form-to-norm estimate for the
difference between the two insertions. Each insertion is bounded
between the energy form domains by \eqref{ch:compressed-form}.
At commuting centers opposite color bases have conjugate determinants.
For eight supercharges a reversal of the root pair interchanges two
even-dimensional occupied wedges, and equivariant radial transport
preserves the resulting normalizer action. For four supercharges the
orientation is already fixed by \eqref{ch:vacuum-phase}.

A local wave packet with disjoint images under the cosets of a finite
stabilizer can be normalized before summing over gauge images.
For \(H\subset K\) the resulting map is
\(\sqrt{[K:H]}\,\Pi_K\) from an \(H\)-invariant seed to \(K\)-invariants,
where \(\Pi_K\) is group averaging. The indices multiply along a chain
of stabilizers, so the maps compose. Starting at the intersection handles
incomparable stabilizers and counts each physical channel once.

\subsection{A common dilation on overlapping charts}

Two charts describing the same separation must use the same physical
dilation where they overlap.
Outside the interaction region the center flats form a finite arrangement,
closed under intersection, with dimension changes in multiples of \(d\).
On a fixed flat the distance to the kernel of a restricted root
\(\chi\) is \(|r_\chi|/\|\chi\|\). The nested cylinders of
Ref.~\cite{Qi}, App.~E.2, can consequently be chosen thinner than every
active root margin. On a section \(W=0\) of an overlapping chart the
central torus fixes \(X\) and its locally unique nearest center in the
other chart. The two center tori therefore commute. Projection onto
a common Cartan recovers both centers by orthogonal projection onto
their center flats. These projections identify the linear arrangement used to
construct the cutoffs even at noncommuting \(U\). At \(W=0\) each transverse root space has nonzero torus weight,
so the first normal derivative of an invariant scalar cutoff vanishes.
On this section the cutoff depends only on the direction of the center,
hence its leading switching current is tangent to the sphere of centers.

In the positive-commutator estimate the generator restricts to dilation
in the free centers and a small positive multiple of the generator on
each surviving simple factor. Taking the rate from rank \(r\) to rank
\(m\) to be \(\prod_{j=m+1}^{r}\epsilon_j\) makes the rate on a common
retained factor independent of the intervening flats. On overlaps the
two free fields differ by \(O(|y|)\) in the normal \(y\) to a collision
flat. Spreading their interpolation over a logarithmic collar of width
\(K\) costs \(O(K^{-1})\) in symmetric derivative, which preserves
strict positivity along the active free direction. When differentiating
both \eqref{ch:scalar} and \eqref{ch:berry} along the dilation, we keep the brackets in the sorted words. Through two such derivatives a
retained bracket costs \(C(1+\sqrt{V_{\rm ret}})\) and a single retained
letter costs \(C(1+|U|)\). The retained projection of the full
bracket is its internal bracket plus \(O(W^2)\), and its transverse linear
part controls \(R|W|\). The bracket identity differentiated along the
matched retained fields gives the same bounds for its first two derivatives as in
Ref.~\cite{Qi}, App.~C.2. The differentiated phase terms thus have
size \(C R^{-3}(1+|U|)(1+\sqrt{V_{\rm ret}})\). They fit the zeroth-order
discrepancy from \eqref{ch:hessian}, while the first-order displacement has
coefficient \(O(R^{-2})\). These bounds control differentiation along
the dilation. The kinetic energy also differentiates transverse to that
field on an overlap of two charts.

Differentiating the transverse quadratic Hamiltonian twice can produce
terms of order \(R\), the oscillator frequency itself. Write \(F_0\)
for that Hamiltonian with its frozen ground energy subtracted, \(J_0\)
for the uncut normalized ground column, and \(A_{\rm src}\) for the
generator on the complete source. If \(A\) is the lifted generator,
set \(E_0=AJ_0-J_0 A_{\rm src}\). Expanding the double commutator and
using \(F_0J_0=0\) on both sides gives the form identity
\begin{equation}
 J_0^* i[i[F_0,A],A]J_0=2E_0^*F_0E_0.                 \label{ch:double}
\end{equation}
Since \(F_0\) annihilates the ground line, only
\(Q_0E_0\) contributes on the right, with \(Q_0=1-J_0J_0^*\).
In particular any scalar phase along that line has zero normal
excitation energy, even though it enters the comparison of actual
waves in \eqref{ch:berry}. The precision and displacement estimates
above control the complementary component, including its
\(O(R^{-2})\) coefficient on a slow derivative. Cutting off the
Gaussian introduces exponentially small tails.

\subsection{Ordinary derivatives across the overlap}\label{app:overlap-jets}
When a slow kinetic derivative is moved off the source in the second
commutator, it may land on a coefficient of the first commutator in any
slow direction. Let \(A_0\) transport the ground state along the common
source velocity in the finest active chart. For either
\(B_{1,0}=i[H_G,A_0]\) or \(B_{1,\Delta}=i[H_G,A-A_0]\), the coupling
from the ground state to the excited transverse states has the form
\[
 Q B Jp=\sum_{|\alpha|\le2} k_{B,\alpha}(s)\,\partial_s^\alpha p,
 \qquad s=(U,z).
\]
For \(j=0,1,2\) let \(M_{j,n}(B)\) be the maximum, over
\(|\alpha|=j,\ |\beta|\le n\), of
\(\|\partial_s^\beta k_{B,\alpha}\|_{f,m}\). The norm is the
operator norm on all retained fermion polarizations with output in the
fast norm \(\|f\|_{f,m}=\|(1+\mathcal N)^{m/2}f\|_{L^2}\).
We use a smooth frame of the
normalized outer shape and a Cartesian frame for the remaining internal
fermions. The reference phase of the occupied line is transported at
fixed outer shape as in the preceding comparison. All indices \(n,m\) are fixed before \(R\) grows. Suppressing
\(n\) in the triple, we will prove
\begin{equation}
 \begin{aligned}
 (M_0,M_1,M_2)(B_{1,0})&\le C_{n,m}(R^{-1/2},R^{-1},R^{-3}),\\
 (M_0,M_1,M_2)(B_{1,\Delta})&\le C_{n,m}(R,1,R^{-2}).
 \end{aligned}                                                \label{ch:jet-condition}
\end{equation}
One ordinary derivative, \(n=1\), suffices for the second physical commutator.

\paragraph{Transport of the transverse ground state.}
On a shell \(R<|z|<2R\), the size of the internal matrices may pass
through the support of a lower-rank cutoff even as the outer separation
increases. We therefore keep the internal matrices in Cartesian
coordinates and hold \(R\) fixed when differentiating in \(s=(U,z)\).
For the physical dilation field \(v\) and its Hermitian action \(M\)
on the finite fermion space we will use
\begin{equation}
 |v(X)|\le C\langle X\rangle,\qquad
 |D^k v(X)|\le C_k\quad(k\ge1),\qquad
 \|D^k M(X)\|\le C_k\quad(k\ge0),
 \label{ch:field-ordinary}
\end{equation}
where \(\langle X\rangle=(1+|X|^2)^{1/2}\) and \(D\) is an ordinary
Cartesian derivative in a fixed physical frame. We establish these
bounds by induction in semisimple rank for every fixed \(k\), since a
Taylor remainder can use one extra derivative of a lower-rank field.
The free-center dilation is linear, its finite action vanishes, and
a smooth modification on a compact set preserves the bounds.

Only the gauge orbit of the separating center enters the choice of
frame. Dividing its longitudinal tangent matrix by \(R\) gives a
uniformly invertible matrix on each buffered patch. Applying the
inverse function theorem on the unit shell and restoring lengths
gives \(D^k\Phi=O(R^{1-k})\) for the length coordinates of the
tubular map and its inverse, and \(D^k U_{\rm fr}=O(R^{-k})\) for
the dimensionless change of frame. A gauge-angle increment is measured
in length units at radius \(R\). Coincident eigenvalues of the
internal matrices are allowed throughout this calculation.

In a product chart the slow velocity \(v_s\) consists of free-center
dilation and the lower-rank velocities at their prescribed rates. The
Gaussian precision \(\Gamma_{\rm f}=R\,\gamma_{\rm f}(s/R)\) and the occupied fermion projector
\(P_f=p_f(s/R)\) depend smoothly on the divided coordinates. The
positive square root defining \(\Gamma_{\rm f}\) and a contour around the negative
spectrum for \(P_f\) have bounded shape derivatives even when
eigenvalues on the same side of the gap coincide. In the section
transported radially from \(U=0\), the fermionic connection one-form
in length coordinates has derivatives of order \(O(R^{-1-k})\).
For the Gaussian the normal velocity is \(LW\), where
\(L=-\tfrac12\Gamma_{\rm f}^{-1}d\Gamma_{\rm f}(v_s)\), since
\(\Gamma_{\rm f}L+L^*\Gamma_{\rm f}=-d\Gamma_{\rm f}(v_s)\) is exactly
the change in width along the slow motion.
The one-form \(\Gamma_{\rm f}^{-1}d\Gamma_{\rm f}\) is \(O(R^{-1})\). A derivative hitting
this form gains a power of \(R^{-1}\), while a derivative hitting
\(v_s\) replaces its \(O(R)\) size by a bounded coefficient.
Consequently, even where the internal cutoff varies on a fixed length
scale, the normal lift obeys
\begin{equation}
 \|L(s)\|\le C,\qquad \|D_s^k L(s)\|\le C_k/R \quad(k\ge1).
 \label{ch:normal-lift}
\end{equation}

Because \(|W|\le cR\), the vector \((v_s,LW)\) obeys
\eqref{ch:field-ordinary}. The finite transport adds the induced
fermionic action of \(i[dP_f(v_s),P_f]\) and the scalar connection
evaluated on \(v_s\) to the retained \(M\). Both additions contract
an \(O(R^{-1})\) one-form with the slow velocity.
On returning to physical coordinates, a derivative on the tangent
map \(D\Phi\) or the frame change supplies the inverse power of \(R\)
needed for an undifferentiated velocity. The angular weights, including
those in an interpolation collar of fixed ratios of radii, have
\(k\)-th derivatives \(O(R^{-k})\). Thus their weighted sum has the
same bounds, and symmetrizing the field with the orbit half-density
includes the divergence of that sum. The construction acts equivariantly
on the full retained matrix space and its fermions. A finite normalizer
average preserves the estimates, completing the induction.

\paragraph{Displacement of the internal matrices.}
On the overlap the gaps resolved by both descriptions are comparable
to \(R\), with constants depending on the fixed flag. The remaining
subsystem retains its own dynamics. On the section \(W=0\) choose the
same terminal gauge frame for both descriptions by factoring a local
representative of the fine orbit into a coarse representative and an
inner representative. Changes of this residual frame transform both
the vector and its induced finite connection. After the linear
identification of the centers, the two slow coordinates then agree
on the section.

Write the internal coordinate change as \(U'=U+\Gamma(s,W)\). Its
linearization in \(W\) intertwines the central torus, whereas the
normal modes carry nonzero central characters and the retained
matrices are neutral. It follows that \(D_W\Gamma(s,0)=0\) for every
\(s\). Taylor expansion normal to the section and differentiation
in its unit-shell coordinates give
\begin{equation}
 |\partial_s^\beta\partial_W^b\Gamma(s,W)|
 \le C_{\beta,b} R^{-1-|\beta|}|W|^{2-b},
 \qquad b=0,1,2 .
 \label{ch:neutral-shift}
\end{equation}
Here \(b\) counts normal derivatives. Higher normal derivatives are
bounded by smoothness of the same unit-shell coordinate change. The
normal coordinate itself transforms as \(W'=B(s)W+O(W^2/R)\) with
\(D_s^k B=O(R^{-k})\). A scalar cylindrical cutoff has normal remainder
\(O(W^2/R^2)\).

The quadratic displacement already occurs in a two-color fragment.
In Hermitian \(SU(3)\) coordinates take
\(X_1=\operatorname{diag}(R,R,-2R)\) and
\(X_2=w(E_{13}+E_{31})\). In the upper two-color space selected by
\(X_1\), the compression of \(X_1+X_2\) is \(R I_2\). In the upper
space selected by the sum itself, the two eigenvalues are \(R\) and
\((\sqrt{9R^2+4w^2}-R)/2\). Half their difference gives a traceless
splitting of \(w^2/(6R)+O(w^4/R^3)\).

A shift of this size can still sample the cutoff of a lower-rank
velocity. For any coefficient \(f\) of the internal field or its
finite transport, we use the identity
\[
 f(U+\Gamma)-f(U)=\int_0^1 Df(U+t\Gamma)\,\Gamma\,dt .
\]
Each term retains either \(\Gamma\) or a derivative of it. On
\(|W|\le1\), Eqs.~\eqref{ch:field-ordinary} and
\eqref{ch:neutral-shift} therefore bound the difference and all its
fixed slow derivatives by \(C_\beta |W|^2/R\). One normal derivative
gives \(C_\beta |W|/R\), and two or more give \(C_{\beta,b}/R\).

The slow component of a velocity pulled back to the common chart is
\((D_{s'}s)v_{s'}+(D_{W'}s)L'W'\). Its value at \(W=0\) agrees after
summing the chart weights because the rates on each common internal
factor agree and the free-center velocities have the prescribed
interpolation. Away from the section, the slow Jacobian and the
weights change by \(O(W^2/R^2)\) against a velocity of size \(O(R)\).
The change of the internal field has the bound just derived, while
the normal velocity gives \(O(|W|/R)O(|W|)\). Their sum has slow
component \(O(W^2/R)\) after subtraction of the common lift. Its
normal component is a bounded linear map on \(W\) plus \(O(W^2/R)\),
by differentiation of the formula for \(W'\). The same orders hold
under any fixed number of ordinary slow derivatives on \(|W|\le1\).

In oscillator coordinates \(y=\sqrt R\,W\), the slow coefficient of
the difference \(\Delta A=A-A_0\) is consequently \(O(R^{-2})\) times a
polynomial in \(\langle y\rangle=(1+|y|^2)^{1/2}\), including fixed
\(s,y\) derivatives. The normal vector, the half-density divergence
and the finite multiplier have order one in these units, since
\(W\partial_W=y\partial_y\). Outside the inner tube
\(|y|>\sqrt R\), the polynomial bounds on the physical coefficients
and their derivatives cost at most further powers of \(\langle y\rangle\).
Every fixed such moment on the vacuum tail is exponentially small in
\(R\). For a general wave in the complement, \eqref{ch:field-ordinary} and the
operator graph norm give the corresponding bound on the full internal
space.
\paragraph{Excitation out of the ground state.}
A slow derivative can create a transverse excitation by changing
the ground state itself or by acting through the metric of the
gauge slice. We separate the two contributions at fixed shell
radius \(R\). In the coordinates \(s,y\), with the unitary dilation
of the fast measure understood, write the slow derivatives on the
right in the Hamiltonian,
\begin{equation}
 H_G=V_{\rm ret}-\tfrac12\Delta_s+
 H_0+\sum_\mu H_{1,\mu}\partial_{s_\mu}
 +\sum_{\mu,\nu}H_{2,\mu\nu}\partial_{s_\mu}\partial_{s_\nu}.
 \label{ch:graded-H}
\end{equation}
The three coefficient operators in \(y\) have orders \(R,R^{-1},R^{-3}\),
respectively, and \(H_{2,\mu\nu}\) is scalar multiplication. Here and
below an order in \(R\) allows a polynomial in \(\langle y\rangle\)
in the coefficient of each normal derivative. Such operators send
a sufficiently high Schwartz seminorm into any prescribed oscillator
moment. The normalized Gaussian obeys these bounds uniformly on the
patch, and its differentiated cutoff errors decay faster than every
inverse power.

Write \(r_{\rm int}=|U|\) for the size of the internal matrices. In the horizontal transverse
coframe the slow inverse metric differs from the identity by
\(O(W^2/R^2)\), the mixed inverse metric is
\(O((r_{\rm int}+|W|)|W|/R^2)\), and the normal inverse metric differs from
its value on the section by \(O(|W|/R)\). For example, a mixed
derivative multiplies the mixed metric by
\(\partial_W=\sqrt R\,\partial_y\). Since \(r_{\rm int}\le CR\) and
\(W=y/\sqrt R\), its coefficient costs \(O(R^{-1})\) in oscillator
moments. The moving frame also adds \(O(R^{-1})y\partial_y\) to a
slow derivative. Two normal derivatives on the normal metric correction
cost \(O(R^{-1/2})\). Differentiating its coefficient in the divergence
form gives the same power after scaling.

Eliminating the longitudinal gauge momentum introduces the fermionic
Gauss generator \(\mathsf G\). Denote the slow, transverse and
longitudinal rows of the omitted gauge orbit by \(C,B,A_\parallel\),
and the longitudinal row of a slow tangent by \(E\). In length
coordinates \(C=O(W)\), \(B=O(r_{\rm int}+|W|)\), \(E=O(W/R)\) and
\(L_\parallel=A_\parallel-EC\) has inverse of order \(R^{-1}\).
The longitudinal covector is
\(L_\parallel^{-t}(\mathsf G-C^t d_s-B^t d_W)\), where the covectors
\(d_s,d_W\) include the orbit half-density. With
\(S=L_\parallel^{-1}(1+EE^t)L_\parallel^{-t}\), the cross terms of
\(\mathsf G\) with slow and normal derivatives in the kinetic energy
have coefficients
\[
 -(E^tL_\parallel^{-t}+CS)\mathsf G
       =O(|W|/R^2),\qquad
 -BS\mathsf G=O((r_{\rm int}+|W|)/R^2).
\]
The normal coefficient costs \(O(R^{-1/2})\) upon application of
\(\partial_W\), and the square of the generator costs \(O(R^{-2})\).
The logarithmic derivative of \(|\det L_\parallel|\) has size
\(O(R^{-1})\) in \(s\) and \(O(|W|/R^2)\) in \(W\). The missing
constant in the latter derivative is the trace of a nonzero-character
variation against a character-preserving inverse at \(W=0\), which
vanishes. In half-density form the two real first-derivative terms
combine into a scalar of order \(R^{-2}\). On the fixed angular
margin all divided matrices used here are smooth in \(s/R,W/R\).
An ordinary derivative of a microscopic coefficient thus gains
\(R^{-1}\), also in its divergence and for repeated restricted roots.

On applying \eqref{ch:graded-H} to \(Jp\), the perpendicular projection
annihilates the frozen ground energy and the retained potentials,
which preserve the fast ground line. Among the remaining multipliers the cubic normal term
\(O(R|W|^3)\) and the linear normal fermion term \(O(W)\) both have
size \(R^{-1/2}\) on the Gaussian, while its quartic normal moment
is \(O(R^{-2})\). In the free slow Laplacian the term with both
derivatives on the source vanishes by \(QJ=0\). A derivative on the
Gaussian or the occupied projector gains \(R^{-1}\) by the smooth
shape dependence of its precision or spectral projection. Hence the
coefficients of \(K=QH_GJ=\sum_\alpha k_\alpha\partial_s^\alpha\)
obey
\begin{equation}
 \|\partial_s^\beta k_\alpha\|_{f,m}\le
 C_{\beta,m} R^{-a_j-|\beta|},\qquad
 j=|\alpha|,\qquad (a_0,a_1,a_2)=(1/2,1,3)
 \label{ch:native-coefficients}
\end{equation}
in every fixed oscillator moment \(m\). The orders in
\eqref{ch:native-coefficients} apply before any internal wave has
been chosen.

The generator \(A_0\) transports the uncut ground state along the
source generator \(a\), so \(A_0J=Ja\) there. Cutting the normal
Gaussian introduces finite-rank fast operators whose kernels and
derivatives are exponentially small. Up to those errors,
\(Q i[H_G,A_0]J=i(Ka-A_{0,Q}K)\), with \(A_{0,Q}\) the complementary
compression. On a term \(k_\alpha\partial_s^\alpha p\), the two
principal transports of \(p\) cancel. What remains is normal linear
and finite matrix transport on \(k_\alpha\), differentiation of that
coefficient along \(v_s\), and \([\partial_s^\alpha,a]\) on \(p\).

For instance, when all \(n\) ordinary derivatives of
\(v_s\cdot\partial_s k_\alpha\) fall on \(k_\alpha\), Eq.~\eqref{ch:native-coefficients}
gives \(O(R^{-a_j-n-1})\) against a velocity of size \(O(R)\). In
every other term a derivative of the velocity is bounded by
\eqref{ch:field-ordinary}. In \([\partial_s^\alpha,a]\) the term with
an undifferentiated velocity cancels, so at most \(j\) slow derivatives
remain on \(p\). Including the half-density divergence, the product
rule through \(n\) further derivatives uses at most \(n+3\) derivatives
of the vector field and \(n+2\) of its finite transport. These bounds
give the first line of \eqref{ch:jet-condition}.

\paragraph{The commutator with the change of lift.}
The lift difference has the form
\(\Delta A=A-A_0=(\Delta A)_0+\sum_\mu z^\mu\partial_{s_\mu}\), where
\((\Delta A)_0=O(1)\) on fast seminorms and \(z^\mu=O(R^{-2})\) is scalar,
with the same bounds after ordinary slow differentiation. Commuting
with the free slow Laplacian in \eqref{ch:graded-H} gives orders
\(1,1,R^{-2}\) at slow derivative orders zero, one and two. The
retained potential contributes \(-z\cdot\partial_s V_{\rm ret}\).
The quartic potential has \(|\partial_s V_{\rm ret}|\le Cr_{\rm int}^3\)
and \(|\partial_s^2 V_{\rm ret}|\le Cr_{\rm int}^2\). The product and its
derivative therefore cost at most \(O(R)\), including at commuting
configurations where \(\sqrt{V_{\rm ret}}\) can fail to be differentiable.

At order zero, \([H_0,\Delta A]\) costs \(O(R)\), while its order-one coefficient
\([H_0,z^\mu]\) costs \(O(R^{-1})\) because the normal derivatives
can act on \(z^\mu\). The terms from \(H_1\partial_s\) have orders
\(R^{-1},R^{-1},R^{-3}\), with the last coefficient coming from
\([H_{1,\mu},z^\nu]\). The second-order slow metric is scalar, so
\([H_{2,\mu\nu},z^\lambda]=0\) removes the only prospective third
slow derivative, leaving coefficients \(O(R^{-3})\) through order two.
Further ordinary slow derivatives preserve these upper bounds by
\eqref{ch:field-ordinary} and the estimates on \(\Delta A\). Projecting onto
\(Q\) after applying the commutator to \(Jp\), including the terms in
which a slow derivative acts on \(J\), gives
the second line of \eqref{ch:jet-condition}.

In the second physical commutator a third slow derivative can occur
when \(\Delta A\) acts on \(QB Jp\), with \(B\) either first commutator.
Its coefficient \(z^\lambda k_{B,\mu\nu}\) and the first ordinary
derivative of that coefficient are both \(O(R^{-4})\). In a pairing
with a complementary wave \(q\), one integration by parts therefore
bounds the contribution by
\[
 C R^{-4}\bigl(\|\nabla_s q\|+\|q\|\bigr)\,\|\nabla_s^2 p\|.
\]
The graph estimate on the complement controls \(\nabla_s q\), while
on the source the weighted Hessian estimate controls
\(R^{-1}\nabla_s^2p\). Both bounds hold in operator norm on the retained
fermion space, including for an unbound internal wave entangled with a
free center.
\paragraph{The vacuum displacement and the common source domain.}
The \(O(R)\) entry of \eqref{ch:jet-condition} is used only for
the mixed graph pairing in the second commutator. The first
commutator on the energy form uses a sharper cancellation.
For the uncut ground state the vacuum diagonal of \(i[F_0,\Delta A]\) is zero,
and its outgoing column is \(iF_0Q(\Delta A)J_0\).
By \eqref{ch:hessian} and the directional bounds preceding
\eqref{ch:double}, the zeroth coefficient of \(Q(\Delta A)J_0\) has size
\(O(R^{-3/2}+(1+r_{\rm int})(1+\sqrt{V_{\rm ret}})/R^3)\), where
\(r_{\rm int}=|U|\). Its first slow differential coefficient has
size \(O(R^{-2})\) by \eqref{ch:neutral-shift}. Applying \(F_0^{1/2}\) costs \(O(\sqrt R)\) in these fixed oscillator moments.
The divided source gradient and potential bounds \eqref{ch:form}
then give the mixed first-form estimate on an arbitrary internal wave.

On the operator domain we also need an ordinary derivative of the
relative lift. In their own frames the two linear normal velocities
have coefficients with \(D_sL=O(R^{-1})\) by \eqref{ch:normal-lift}.
The change to the common normal frame has first and second derivatives
of orders \(R^{-1}\) and \(R^{-2}\), respectively. Differentiating
the induced correction from motion of the frame costs \(O(R^{-1})\), since
\(v_s=O(R)\) and \(D_sv_s=O(1)\). Subtraction now bounds the
ordinary derivative of the difference of the two linear lifts by
\(C/R\). Multiplication by a fixed normalized Gaussian polynomial
preserves the estimate. This use of smoothness in the outer shape is
confined to the differentiated graph estimate. The undifferentiated
difference and its derivatives along the flow retain the bracket gain
of \eqref{ch:hessian}--\eqref{ch:berry} in the energy form.

The nonlinear change contributes \(O(R^{-3/2})\) after one normal
derivative of the Gaussian. On the common section the retained finite
matrix multipliers agree. Evaluating them off that section introduces
the slow displacement \eqref{ch:neutral-shift} and a relative frame
rotation \(O(|W|/R)\). The relative scalar phase has
\(D_sm=O(R^{-1})\) and \(D_s^2m=O(R^{-2})\) by outer-shape smoothness.
Its contraction with \(v_s\) therefore also has gradient \(O(R^{-1})\),
including where \(D_sv_s\) differentiates a fixed-size internal cutoff.
These ordinary derivative bounds apply on either side of the mixed
pairing in Ref.~\cite{Qi}, Apps.~C.3--C.4. For a complementary state
carrying internal color we use the scalar comparison preceding
\eqref{ch:mixed-form}.

Compact transfer requires the projected first commutator to decay at large
radius even if the phase of the two lifts differs by order one. Set
\(\zeta=J^*(\Delta A)J\), \(E_Q=Q(\Delta A)J\),
\(D_h=J^*H_GJ-h\) and \(K=QH_GJ\). The two rows of the localized
Hamiltonian give the exact identity
\[
 J^*i[H_G,\Delta A]J=i[h+D_h,\zeta]+i(K^*E_Q-E_Q^*K).
\]
In the first term the slow kinetic energy differentiates the scalar phase.
Its gradient is \(O(R^{-1})\), which leaves \(O(R^{-1/2})\) after
\eqref{ch:form} bounds the source gradient. The mean velocity error is
\(O(R^{-2})\). Its commutator with the retained potential has a term of
size \(C r_{\rm int}\sqrt{V_{\rm ret}}/R^2\) and therefore gains the same
power by \eqref{ch:form}. In the second term the slow metric and velocity multiply the same
vacuum by real scalars, with the common factor \(-i\) in the velocity
operator. The two principal color overlaps therefore agree, and the
apparent third slow derivative cancels. One integration
by parts in the remaining second-order terms allows one divided gradient
on each side, while the lower-order terms obey
\eqref{ch:native-coefficients}. The resulting exterior form bound
has the order in Eq.~(C.26) of Ref.~\cite{Qi} and includes the phase
of the occupied line at the interacting background.

For an independent free spectator, the subsystem identification acts
as the identity on the spectator factor and therefore commutes with
its kinetic energy. In a simultaneous refined chart the resolved
gaps have fixed margins, the extra center enters only the free
coordinates and smooth masks, and the same estimates apply. If the
spectator alone is resolved on its scale, the interacting factor is
kept whole and its own coefficient estimates are measured at its own
radius. Its transverse gap is measured by the internal separation alone.
The product with a spectator therefore obeys the same mixed bounds. The first commutator is
bounded on the physical energy form domain and the second as a form
on two operator-domain vectors by the ordered graph comparisons of
Ref.~\cite{Qi}, Apps.~B--D. Sandwiching the latter by the nonreal
resolvent gives its required second commutator bound.

\subsection{Propagation of the complete source}

At semisimple rank one every proper centralizer is abelian, so the source
commutator is \(2T_z\), twice the center kinetic energy. At higher rank
the threshold energies are sums of lower-rank exceptional energies.
On a sufficiently small joint spectral cell away from those sums, either
the center has positive kinetic energy or one internal factor lies in a
regular interval. The other factors admit an arbitrarily small negative
lower bound on the cell, even at an eigenvalue, because the virial
identity makes the compression of the commutator to its eigenspace zero.
The lower bounds are scalar functions of the other factors' spectral
parameters and can be added for an entangled source as well as a product
state, as in Ref.~\cite{Qi}, Apps.~F.2--F.3.
The transfer to the physical Hamiltonian uses two consequences of the
local comparison. Write \(E_I\) for the physical spectral projection
on a compact interval \(I\), \(j\) for a cutoff buffered inside a
chart, and \(\eta\) for a smooth function equal to one on a neighborhood
of \(I\). The estimates, in the indicated orders, are
\begin{equation}
 \begin{aligned}
 \|(H_Q+C)^{1/2} R^{1/4}QjE_I\|&\le C_I,\\
 \|(h+C)^{1/2}R^{1/8}(1-\eta(h))J^*jE_I\|&\le C_{I,\eta}.
 \end{aligned}                                           \label{ch:radial-moments}
\end{equation}
For the first inequality the complementary graph norm controls the
\(R\)-weighted gradient square and the squares of the oscillator and
retained potentials. A smooth truncation \(w\) of \(R^{1/4}\) has
\(w^2\le CR\), \(|\nabla w|\le CR^{-3/4}\), and
\(w^2V_{\rm ret}\le C(R^2+V_{\rm ret}^2)\). The complementary form
of \(wq\) is consequently bounded by the graph norm squared of \(q\),
uniformly as the truncation is removed. For the second inequality apply the source row to
\(p=J^*jE_I\) and \(q=QjE_I\). With \(H_I=H_G|_{\operatorname{Ran} E_I}\),
the difference \(hp-pH_I\) consists of the error on the ground line,
the adjoint coupling on \(q\), and \(J^*[H_G,j]E_I\).
The first two terms obey \eqref{ch:compressed-form} and
\eqref{ch:mixed-form}, and the spatial cutoff costs \(C/R\) on one
derivative. In the two spectral variables the inverse on
\(1-\eta(h)\) has kernel \((1-\eta(e))/(e-\lambda)\), \(\lambda\in I\).
Its denominator stays away from zero and gains the source form scale
at large \(e\). Commuting a truncated \(R^{1/8}\) through this
inverse costs \(O(R^{-7/8})\) on one form derivative. The
\(R^{1/4}\) moment on \(q\) absorbs the weight in the coupling term,
which proves the second inequality by the argument of Eq.~(E.1)
of Ref.~\cite{Qi}.

Both components now have small tails in their respective form norms.
For compactness on a bounded region we localize the source before
applying the spectral filter, since commuting a smooth radial cutoff at
radius \(L\) through \(\eta(h)\) costs \(O(L^{-1})\) in form norm.
Local ellipticity gives compactness from graph to form on the
localized part. In the squared partition the leading switching
currents sum to zero because \(\sum_Z j_Z\nabla j_Z=0\), and the remaining
terms contain a decaying derivative of a cutoff or the difference
of the two lifts estimated above. The physical commutator thus differs
from the sum of source commutators by a compact operator on the energy
window, as in Eq.~(F.2) of Ref.~\cite{Qi}.

At each rank the thresholds are a finite union of sums of lower-rank
closed countable exceptional sets. Since the energies are nonnegative,
each summand in a bounded sum is bounded. The sums are therefore
closed. Away from the thresholds the Mourre estimate makes the
eigenvalues locally finite with finite multiplicity. Adjoining the
point spectrum of \(H_G\) to the thresholds gives a closed countable set
\(\mathcal F_G\). The strict Mourre estimate on its complement and
the two bounds on physical commutators give the limiting absorption
estimate of Gol\'enia and Jecko~\cite{GoleniaJecko} with weights in the
generator \(A\). In particular singular
continuous measure would have to be supported on the countable set
\(\mathcal F_G\), where every continuous measure vanishes.

The physical radius \(\sigma=(1+|X|^2)^{1/2}\) is the weight needed
both for escape and for testing the resolvent on wave packets.
On the ground line, division of the dilation by \(R\) leaves a
bounded coefficient on the free center momentum. For a retained
factor with radius \(\sigma_a\), write
\(R^{-1}A_a=(\sigma_a/R)(\sigma_a^{-1}A_a)\) and apply the lower-rank
form estimate to the second factor. The error between \(AJ\) and
\(JA_{\rm src}\), divided by \(R\), is form-bounded by the displacement
estimate above. The complementary form controls the physical derivative
on \(Q\),
and the derivatives of the chart cutoffs remain bounded after division
by \(\sigma\). Hence
\(\|\sigma^{-1}Au\|\leq C\|u\|_{\mathcal Q(H_G)}\).
For a real smooth cutoff \(f\) on a regular band, take the adjoint of
\(\sigma^{-1}Af(H_G)\) and use the bounded commutator
\([A,f(H_G)]\) to obtain the bounded operator \(Af(H_G)\sigma^{-1}\).
Interpolation gives \(\langle A\rangle^b f(H_G)\sigma^{-b}\)
bounded for \(0<b<1\). Applying this map and its adjoint on the two
sides of the conjugate-weight resolvent, with \(f=1\) around
\(I\Subset\mathbb R\setminus\mathcal F_G\), gives
\begin{equation}
 \sup_{\substack{\operatorname{Re}z\in I\\0<|\operatorname{Im}z|\le1}}
 \big\|\sigma^{-b}(H_G-z)^{-1}\sigma^{-b}\big\|<\infty ,
 \qquad \tfrac12<b<1 .                                   \label{ch:radius-lap}
\end{equation}
For the complementary part of the resolvent the distance from \(I\) to
the omitted spectral support gives a uniform bound. This change of
weight is the one in Ref.~\cite{Qi}, App.~F.
The minimal-velocity bound for \(A\) and the ordered estimate on
\(Af(H_G)\sigma^{-1}\) give an \(O(\delta)\) upper bound on the limiting
norm in \(\sigma<\delta |t|\). At the outer edge, apply the nonnegative
energy form to \(e^{i\lambda\phi}u\) for a capped radial function
\(\phi\). Its quadratic dependence on real \(\lambda\) bounds the outward
current by \(\sqrt{2q_{H_G}[u]}\|\nabla\phi\,u\|\), so the radius
expectation grows at most linearly for data of finite first moment.
Truncation and density give arbitrarily small limiting norm in
\(\sigma>M|t|\) as \(M\) increases. The inner and outer velocity bounds
are the versions of Eqs.~(4.23) and (4.25) of Ref.~\cite{Qi} for \(H_G\).

\paragraph{Angular flux on the two evolutions.}

The tangential current from a smooth angular cutoff \(c\) factors on
\(D=d\dim Z\) free coordinates as
\begin{equation}
 \begin{aligned}
 i[-\Delta_z/2,c]&=B^*K_cB, &
 B&=|z|^{-1}\Lambda^{1/2},\\
 K_c&=\Lambda^{-1/2}i[-\Delta_{S^{D-1}}/2,c]\Lambda^{-1/2},&
 \Lambda&=\sqrt{-\Delta_{S^{D-1}}+(D-2)^2/4}.
 \end{aligned}
 \label{ch:angular}
\end{equation}
The angular commutator is first order, so the two half-derivatives in \(K_c\) make it bounded. In the three-dimensional scalar partial
wave the eigenvalue \(\lambda\) of \(\Lambda\) is \(1/2\). The bounded
dilation phase used in Ref.~\cite{Qi} has to allow conjugation by one
power of \(|z|\), which requires a wider strip of analyticity in this
partial wave. In \(s=\log|z|\) with its Lebesgue half-density, put
\(D_s=-i\partial_s\), \(b=\max(\lambda,3/2)\) and
\(A_{\rm ph}=\arctan(D_s/b)\). The radial kinetic energy is
\(T_\lambda=e^{-s}(D_s^2+\lambda^2)e^{-s}/2\). Shifting its Mellin
multiplier by \(i\) and \(-i\) gives
\begin{equation}
 i[T_\lambda,A_{\rm ph}]=e^{-s} C_{\lambda,b}(D_s)e^{-s},\qquad
 C_{\lambda,b}(\xi)=\frac{\xi^2+\lambda^2}{4}
       \log\frac{\xi^2+(b+1)^2}{\xi^2+(b-1)^2}
       \geq \tfrac{3}{13}\lambda .                      \label{ch:mellin}
\end{equation}
The inequality follows from \(\operatorname{arctanh}x\ge x\) at
\(x=2b/(\xi^2+b^2+1)\). The shifted kernels are uniformly bounded since
\(b\ge3/2\). Integration in time gives a bound for
\(\int\|B e^{-itT_z}u\|^2\,dt\), valid after tensoring with an arbitrary
independent internal evolution. The positive-dilation estimate in the
moving annulus also persists because \(b\le3\lambda\).

The current between the two evolutions requires a time-integrated bound
on each state. On the physical side, the equation for
\(p_Z(t)=J_Z^*j_Ze^{-itH_G}f(H_G)\psi\) contains both the intrinsic
comparison error and the flux of the chart cutoff \(j_Z\).
Here \(f\) selects a regular compact energy interval and \(J_Z\) inserts
the normal ground state. In the radial annulus the localization error is
paired using the square-integrable mismatch
\((P_z-z/t)p_Z(t)\) in measure \(dt/t\), where \(P_z=-i\nabla_z\). The angular current is evaluated
in the chart of its own cutoff so the bounded radial phase need not
commute with a change of angular chart. On the source side the internal matrices may also separate. We apply the
same forced-equation estimate to that subsystem with one extra free
coordinate \(v\in\mathbb R^d\) representing the distant fragment.
The insertion on the product is \(J\otimes I_v\). The free angular bound tensors with the internal identity. Upon
diagonalizing the internal Hamiltonian, the free estimate is uniform in
its nonnegative spectral parameter, including at an internal threshold.
The coupling and cutoff terms in the projected equation are controlled
by the same weighted bounds \eqref{ch:radial-moments} for the subsystem
with the extra free coordinate.
During simultaneous breakup, on a common annulus some component
\(v_j\) of the original exterior coordinate is comparable with the
refined radius. For distinct indices \(a,b\) in the other free
coordinates, the rotation generators obey
\[
 \Omega_{ab}=(z_a/v_j)\Omega_{jb}-(z_b/v_j)\Omega_{ja}.
\]
Their coefficients and angular derivatives are bounded on a smaller
patch. Interpolating the localized bounds at zero and one angular
derivative gives the half derivative in \eqref{ch:angular} from the
subsystem estimates with a free spectator. Applying the induction with
that spectator present, as in Ref.~\cite{Qi}, Apps.~E.4--E.5, gives the
angular bound on the comparison evolution before we construct the inverse
wave operator.

\subsection{Recovering the escaping probability}

On an annulus of positive speed we project the evolving physical state
onto the fast vacuum, restrict its center momentum to a compact region
away from zero and the collision directions, and evolve the result
backward under \(h\). Denote this comparison by \(L_Z(t)\). In its
time derivative, the angular current \eqref{ch:angular} contributes a
matrix element bounded on \([t_0,t_1]\) by the product of the two
independent \(L^2(dt)\) norms on that interval. For a fixed physical
wave the integral of its squared norm on the tail tends to zero as
\(t_0\to\infty\), uniformly on the source unit ball. Fixing the source
wave gives the corresponding estimate uniformly on the physical unit
ball.

In the intrinsic interaction the bare \(O(R^{-1/2})\) excitation
is paired with the physical complement. The transverse gap supplies \(R^{-1/2}\) in
passing from the complementary form to the norm, and the first
inequality in \eqref{ch:radial-moments} supplies a further \(R^{-1/4}\).
The resulting \(O(|t|^{-5/4})\) bound is integrable on the moving annulus. The energy cutoff error
uses the second moment in \eqref{ch:radial-moments}, with the remaining
source and moving-cutoff errors estimated as in Ref.~\cite{Qi},
Secs.~5.1--5.2. Thus both \(L_Z(t)\) and \(L_Z(t)^*\) have strong
limits.

For an internal eigenvector \(\phi\) of energy \(e\), take a real tail
cutoff \(\eta_R\) supported outside radius \(R\), with
\(\|\nabla\eta_R\|_\infty=O(R^{-1})\). Its eigenvalue equation gives
\[
 q_{H_{\rm int}+C}[\eta_R\phi]
   =(e+C)\|\eta_R\phi\|^2+\tfrac12\|(\nabla\eta_R)\phi\|^2
   \longrightarrow0 .
\]
The form tail therefore vanishes for each internal eigenstate, including
an eigenstate embedded in a continuum. Distinct final channels have disjoint center-velocity
support on dense seed cells or require a bound internal state to have
a tail at macroscopic separation. Their isometric ranges are therefore
orthogonal. On the rest of a complete source, lower-rank completeness
resolves each internal continuum into its final fragments.
Equations~\eqref{ch:hessian}--\eqref{ch:berry} identify their successive
insertions with the direct insertion. The limit of \(L_Z(t)^*\) thus
takes values in the closed sum of bound-fragment ranges.

Suppose a continuous physical state in a regular compact energy interval
is orthogonal to those ranges. Its projected component on each compact
range of intermediate center momenta tends to zero. In an annulus
\(c<|z|/t<M\) the part with \(|P_z|<c/4\) has vanishing logarithmic mean
square by the velocity-mismatch estimate. For \(|P_z|>L\) the squared
norm is \(O(L^{-2})\) by the uniform source form bound and
\(h\ge P_z^2/2\). Taking the time mean before \(L\to\infty\) leaves
zero mean probability in every such annulus. The complementary
transverse space costs \(c'R\) and the region away from the valleys
costs order \(R^4\), so this conclusion holds in physical space after
summing the finite partition. The independent lower and upper radial
escape estimates on the regular interval place the norm in those same
annuli as \(c\downarrow0\) and \(M\uparrow\infty\). The physical state
must therefore vanish. Repeating the argument for the opposite direction of time and taking the closure
over regular intervals gives both wave-operator ranges. Finally the
insertions intertwine fermion number, so restriction
to any fixed occupation commutes with this decomposition.
\section{From an isolated valley to the outgoing flux}
\label{app:transmission}

Consider the pair of spinor valleys of \(\Spin(2n)\) at fixed even \(n\ge4\).
In this appendix \(H=H_{\Spin(2n)}\) is the four-supercharge Hamiltonian
in the gauge-singlet sector with \(F=n(n-1)\) and \(J=0\), except when
we explicitly work in the Cartesian space before projection. Set
\(N=n(2n-1)\), the dimension of the color algebra, and
\(d_X=3N\), the dimension of the Cartesian configuration space.
At \(n=4\), a diagram automorphism carries either spinor valley to the
vector valley, so the same argument applies to every ordered pair of
distinct valleys. Fix a compact interval \(I\) in the first local band
\((e_0,e_1)\). We prescribe a radial derivative in one valley and follow
the resulting solution from negative energy into \(I\). The angular gap
fixes which solution continues across the threshold, while its incoming
and outgoing radial coefficients fix the normalization of the measured
flux. All estimates are at fixed \(n\) and may depend on \(I\).

\subsection{The angular gap in one valley}

The two valley orbits are separated on the unit sphere by the angle
\(\arccos(1-2/n)\). The absolute inner product of unit spinor
directions of opposite parity is largest when their signs differ in
one position. Allowing arbitrary gauge conjugations gives the same
maximum, since stationarity of the color inner product forces the
two representatives to commute, and the centralizer of one then
conjugates the other into a common Cartan algebra. Spatial rotations
allow either sign of the inner product.
Choose disjoint smooth invariant neighborhoods \(\Omega_i,\Omega_j\)
whose closures lie inside the respective centralizer charts and which
the diagram involution exchanges. In each exterior cone
\(D_i=(R_0,\infty)\times\Omega_i\) impose Dirichlet conditions on the
cap and sides. With \(\rho=|X|\), multiplication by \(\rho^{(d_X-1)/2}\)
gives measure \(d\rho\,d\omega\) and the exact radial decomposition
\begin{equation}
 H_{D_i}=-\tfrac12\partial_\rho^2+A(\rho),\qquad
 A(\rho)=\rho^{-2}K_{\Omega_i}+\rho^4 V(\omega)+\rho Y(\omega),
 \label{tx:polar}
\end{equation}
where \(K_{\Omega_i}=-\Delta_{\Omega_i}/2+(d_X-1)(d_X-3)/8\) has
Dirichlet boundary condition. The commutator potential \(V\ge0\) and the
bounded Hermitian Yukawa multiplier \(Y\) are evaluated at unit radius.
Given \(\sup I<m<e_1\), we claim that for sufficiently large \(R_0\)
there is a real normalized angular section \(p(\rho)\) such that
\begin{align}
 \|(A-e_0)p\|&\le C\rho^{-3/2},&
 \|\partial_\rho p\|&\le C\rho^{-1},&
 \|\partial_\rho^2p\|&\le C\rho^{-2},\nonumber\\
 \langle \xi,A(\rho)\xi\rangle&\ge m\|\xi\|^2
 \qquad\bigl(\xi\perp p(\rho)\bigr).
 \label{tx:angular}
\end{align}
The last bound concerns every perpendicular angular form vector and
requires a separate estimate from the construction of \(p\).

\paragraph{Restricting the ground mode to the physical sphere.}
Let \(r\) be the length of the mass-normalized relative center, \(U\) the
three internal matrices, and \(y=\sqrt r\,W\) the rescaled normal
coordinates. Insert the positive internal ground state \(\psi_0(U)\)
and the transverse vacuum at the retained background with constant reduced
radial amplitude. In the component along the vacuum, the error in the
slow inverse metric is \(O(r^{-3})\), while derivatives of the vacuum
and the orbit density contribute \(O(r^{-2})\) and the residual
oscillator energy is bounded by \(Cr^{-3}V_{\rm ret}(U)\) by
\eqref{ch:scalar}. Acting on \(\psi_0\), these terms give a defect of
order \(r^{-2}\). In the perpendicular component the coefficients of
zero, one and two slow derivatives have orders \(r^{-1/2},r^{-1},r^{-3}\)
by \eqref{ch:native-coefficients}, giving a total perpendicular
defect \(g_\perp=O(r^{-1/2})\).

Write \(F_\perp\) for the frozen transverse oscillator measured from
its vacuum energy on the perpendicular space. Its gap is at least \(cr\)
on a sufficiently small fixed chart. Subtracting \(F_\perp^{-1}g_\perp\)
from the inserted state cancels the larger defect at a cost
\(O(r^{-3/2})\) in the wave. The only nonzero central grades on either spinor valley are \(+1,-1\),
which excludes a scalar cubic in the normal coordinates. The slow derivatives and internal operators left
over after the cancellation act with polynomial internal weights and
bounded rescaled oscillator moments. The needed moments of \(\psi_0\)
follow by testing its eigenfunction equation with a truncated polynomial
weight \(w\), which gives
\[
 q_{H_b}[w\psi_0]=e_0\|w\psi_0\|^2+
                         \tfrac12\|(\nabla w)\psi_0\|^2 .
\]
For the present \(SU(n)\) fragment the strength of confinement in
\eqref{eq:bosbound} can be read off from its roots. Normalize the
invariant metric on \(\mathfrak{so}(2n)\) by
\(\|X\|^2=-\tfrac12\Tr_{\R^{2n}}(X^2)\) in the vector representation.
A skew block with eigenvalues \(\pm i x_a\) contributes \(x_a^2\),
so the induced norm on the \(\mathfrak{su}(n)\) Cartan is
\(\|x\|^2=\sum_a x_a^2\). The trace
\(t(x)=\sum_{a,b}|x_a-x_b|\) is linear on the ordered traceless
chamber. Its extremal ray \(v_j\) has \(j\) entries \(-(n-j)/n\)
and \(n-j\) entries \(j/n\), so
\(t(v_j)/\|v_j\|=2\sqrt{nj(n-j)}\). Summing the chamber rays with
nonnegative coefficients and using the triangle inequality gives
\(t(x)\ge2\sqrt{n(n-1)}\,\|x\|\), with equality for \(j=1,n-1\).
Thus at unit coupling the linear term in \eqref{eq:bosbound}
dominates \(\sqrt{n(n-1)/2}\sum_\mu\|U_\mu\|\).
This confining term increases the available
polynomial moment at each application, and the reserved kinetic energy
controls one derivative. Local elliptic estimates on translated unit
balls then give every fixed weighted derivative, since the potential
and its derivatives grow polynomially. In the normal variables the
oscillator inverse preserves the required polynomial Gaussian moments.
Differentiating the divided oscillator and its inverse at fixed \(U,y\)
bounds the residual and one radial derivative by \(Cr^{-3/2}\), and
the corrected wave has bounded moments through two dilation derivatives.

In these coordinates the physical sphere satisfies
\(\rho^2=r^2+|U|^2+|y|^2/r\). Smooth cutoffs to
\(|U|,|y|\le2r^{1/4}\) cost arbitrarily high inverse powers of \(r\)
and give \(0\le\rho-r\le Cr^{-1/2}\). The one-dimensional Sobolev
bound at fixed \(U,y\) therefore controls a residual \(f\) on the graph by
\[
 \|f(r(\rho,U,y),U,y)\|^2
 \le C\int_{\rho-2}^{\rho+1}
       \bigl(\|f(t)\|^2+\|\partial_t f(t)\|^2\bigr)\,dt
\]
in the corresponding half-densities. At fixed Cartesian angle dilation
acts on \(r,U,y\) with weights \(1,1,3/2\). Each physical radial
derivative thus gains one inverse power of \(\rho\), up to the second
derivative needed in \eqref{tx:polar}. The change of radial density
and the moments of \(\rho-r\) give angular norm \(1+O(\rho^{-1})\).
Normalizing gives the first line of \eqref{tx:angular}. The same
moments give an \(O_I(\rho^{-1})\) angular error in replacing the
physical ground-channel insertion of \(e^{\pm ikr}\) by
\(e^{\pm ik\rho}p(\rho)\), for \(E=e_0+k^2/2\in I\).

\paragraph{Excluding a second angular mode.}
Let \(J_0\) insert the uncorrected transverse vacuum and the internal
ground state on reduced radial functions. On a shell \(\rho\asymp R\)
inside the cone, complete the square in the mixed form
\eqref{ch:mixed-form}, reserving half of the transverse excitation gap.
On the vacuum, \eqref{ch:compressed-form} leaves the nonnegative center
kinetic energy and the internal Hamiltonian, which satisfies
\(H_b\ge e_1-(e_1-e_0)|\psi_0\rangle\langle\psi_0|\) on singlets
of all spatial spins. On the perpendicular space the remaining gap is
at least \(cR-C\). Together the two inequalities give
\begin{equation}
 q_H[w]\ge(e_1-o(1))\|w\|^2-(e_1-e_0)\|J_0^*w\|^2 .
 \label{tx:exteriorform}
\end{equation}
On a fixed angular collar separated from the valley orbit, either
\(|U|\ge aR\) or \(|W|\ge aR\) for some \(a>0\). In the first region
\eqref{eq:bosbound} gives energy of order \(R\) on the vacuum, while
in the second its Gaussian overlap is \(O(e^{-cR^3})\). The excited
component costs order \(R\) on both regions, and partitioning between
them costs \(O(R^{-2})\). The lower form bound on the collar
consequently tends to infinity.

For a unit \(\xi\perp p(R)\) with angular expectation bounded above,
test \eqref{tx:exteriorform} on \(w=b_L(\rho-R)\xi\), where \(b_L\)
is a normalized smooth even bump of width \(L=R^{1/3}\). The two
positive terms in \(A(R)\) have total expectation \(O(R)\) because
\(RY\ge-CR\). Averaging over the even bump cancels the linear
variation of each coefficient, so the change in their expectations
and the radial kinetic energy together cost \(O(L^2/R+L^{-2})=o(1)\).

It remains to see why \(J_0^*w\) is small when the two radii
\(\rho\) and \(r\) differ. Put \(\tau(\omega)=r/\rho\), and let
\(p_0\) be the angular wave before the oscillator correction. In
\(dr\,d\omega\) the extraction kernel is
\(\tau^{-1}b_L(r/\tau-R)p_0(r/\tau,\omega)\). On the truncated support,
integrate \(\partial_t p_0(t,\omega)\) from \(R\) to \(r/\tau\)
inside a shell of width \(O(L)\). The angular norm of the integral
is \(O(L/R)\) by Minkowski's inequality, also when the endpoint
depends on \(\omega\). The factor \(b_L=O(L^{-1/2})\) and an
\(r\)-interval of length \(O(L)\) give the same bound in the joint
kernel norm. Here \(|r/\tau-r|=O(R^{-1/2})\), so translating the
bump costs \(O(R^{-1/2}/L)\) in that norm. The Jacobian tends to one,
and changing variables \(r=\tau\rho\) controls the discarded tails
by the moments of the uncorrected wave. Since
\(\|p_0(R)-p(R)\|=o(1)\), the kernel differs by \(o(1)\) from
\(b_L(r-R)p(R,\omega)\), which annihilates \(\xi\).
Equation~\eqref{tx:exteriorform} now gives angular expectation
at least \(e_1-o(1)\) uniformly on the perpendicular space, proving
\eqref{tx:angular}.

\subsection{Continuation through the internal threshold}

On each cone write the regular solution of \((H_{D_i}-z)\phi=0\) as
\(\phi=pa+q_\perp\), where \(Qq_\perp=q_\perp\) for
\(Q=1-|p\rangle\langle p|\). Both components vanish at the cap and we
fix the normalization by \(a'(R_0)=1\). The angular state is real and
normalized, so \(\langle p,p'\rangle=0\). The scalar part of the
radial operator is \(-\partial_\rho^2/2+e_0+v\), with
\(v=\langle p,(A-e_0)p\rangle+\|p'\|^2/2\). Its two off-diagonal
terms are
\begin{equation}
 Ba=-p'a'+u a,\qquad u=Q(Ap-p''/2),\qquad
 B^*q_\perp=(\langle p',q_\perp\rangle)'
                      +\langle u,q_\perp\rangle .
 \label{tx:coupling}
\end{equation}
Both \(u\) and \(v\) are \(O(\rho^{-3/2})\). The slower term \(p'a'\)
excites a perpendicular mode, whose contribution back to the
propagating wave therefore requires a second coupling.

At \(z=-s<0\) a regular radial wave can grow exponentially, while
inside \(I\) its open component oscillates. The gap in
\eqref{tx:angular} leaves a range of exponential weights that selects
the same solution in both regimes. Write \(x=\rho-R_0\),
\(k^2=2(z-e_0)\), and choose \(\gamma>0\) locally in \(z\) with
\[
 |\operatorname{Im}k|<\gamma,\qquad \operatorname{Re}z+\gamma^2/2<m .
\]
For real \(z<e_0\) the available interval is nonempty because
\(|\operatorname{Im}k|^2/2=e_0-z<m-z\).
The free radial inverse with zero value and slope at the cap has kernel
\(-2\,1_{t<\rho}\sin(k(\rho-t))/k\), which is analytic in \(z\)
through \(k=0\). After conjugation by \(e^{-\gamma x}\) this kernel
and its first derivative are integrable. On perpendicular sections
the Dirichlet form controls \(\|q_\perp'\|^2/2+m\|q_\perp\|^2\), and
the weight shifts its real part by \(-\gamma^2\|q_\perp\|^2/2\).
In the resulting first-order norms the couplings in \eqref{tx:coupling}
are \(O(R_0^{-1})\), with the differentiated pairing in \(B^*\) acting
on the scalar inverse by integration by parts. The coupled inverse thus
has a convergent Neumann series for sufficiently large \(R_0\).

The solution \(\phi_i(z)\) with unit slope is holomorphic in the weighted
space. A finite cover of \([-s,\sup I]\) gives a common \(R_0\), and
uniqueness in the larger weighted space identifies the solutions on
overlapping neighborhoods. If the slope is free, the same inverse
solves for arbitrary forcing and leaves a one-dimensional kernel.
We will use this last property to determine the sign of the solution at
\(-s\).

Although \(\phi_i\) can grow along the valley, the angular sides of
the cone lie in the collar where the energy tends to infinity.
In the localized eigenfunction identity use a truncated exponent
\(-(\gamma+\delta)\rho+\min\{M,\rho b(\omega)\}\), where \(b\)
vanishes on the derivatives of the inner cutoff and is a large
constant on the outer collar. The Cartesian gradient of the exponent
is bounded independently of \(M\). At sufficiently large radius the
collar form absorbs its squared gradient, and \(M\to\infty\) gives
decay at any specified exponential rate on the outer collar. Local
elliptic estimates give the same decay for derivatives on a smaller
collar, uniformly on compact \(z\)-sets. At the intersection of cap
and side the faces meet orthogonally. In \((\log\rho,\omega)\)
coordinates the principal part is a product after scalar rescaling,
so Dirichlet reflection across the cap gives second-order regularity
up to the edge. In particular the normal derivatives on both faces
define boundary currents.

\subsection{The sign of the coupling at negative energy}

Set \(\ell=F/2=\binom n2\). In the real adjoint color frame
the one-body Yukawa matrix \(M(X)=\sum_a \sigma_a\otimes i\ad X_a\)
commutes with \(\Theta=(i\sigma_2\otimes1)\mathsf K\), where
\(\mathsf K\) denotes coordinate complex conjugation. Because
\(\Theta^2=-1\), the induced antilinear map
\(\mathcal J=\Lambda^{2\ell}\Theta\) is an involution. It fixes the
paired volume of every quaternionic \(\ell\)-plane
\(P\subset\mathbb H^N\),
\[
 \Phi(P)=\bigwedge_{\alpha=1}^{\ell}
              (v_\alpha\wedge\Theta v_\alpha),
\]
since \(\Theta v_\alpha\wedge(-v_\alpha)
=v_\alpha\wedge\Theta v_\alpha\).
Let \(C\) be the cone of positive sums of these volumes in the real
space \((\Lambda^{2\ell}\mathbb C^{2N})^{\mathcal J}\).
The inner product of two generators is the nonnegative complex
determinant of their quaternionic overlap. The \(\ell\)-th power of
the invariant symplectic form pairs equally and positively with all
unit generators, so their compact convex hull is separated from zero
and \(C\) is closed and pointed. Its real span is invariant under
\(\mathrm{GL}(N,\mathbb H)\), whose complexified Lie algebra
\(\mathfrak{gl}(2N,\mathbb C)\) acts irreducibly on the exterior
power. Hence the generators span the stated real space. For \(C\)
and its dual \(C^*\), finite dimensionality then gives a constant
\(C_0\) with \(\|v\|\le C_0\|w\|\) whenever \(0\le v\le w\)
in the corresponding cone order.

\paragraph{Strict propagation between occupied planes.}
Along a Cartesian Brownian bridge the ordered exponential of
\(-M(X(t))\) is quaternionic linear. Between paired volumes its
matrix element is the complex determinant of the compression between
the two planes, equal to the product of its squared quaternionic
singular values. Each history therefore preserves \(C\) and receives
a positive scalar-potential weight. Heat evolution preserves \(C\),
also with Dirichlet killing, and by self-adjointness it preserves
\(C^*\). Gauge and spatial rotations preserve both cones.

Strict positivity requires full-rank overlap on a set of paths of
positive measure. Put \(L_X=\ad X\) in the real color representation.
For distinct spin axes the constant controls satisfy
\[
 [i\sigma_a\otimes L,i\sigma_b\otimes L']
 =-i\sum_c\epsilon_{abc}\sigma_c\otimes(LL'+L'L).
\]
Repeated brackets with \(i\sigma_a\otimes L_X\), choosing a different
axis from the current one at each step, give \(i\sigma_c\otimes L_X^h\)
for every \(h\ge1\) on every spin axis.
The adjoint Casimir supplies the color identity. Polarizing the
powers gives all symmetrized words, which span the image of the
enveloping algebra by Poincar\'e--Birkhoff--Witt symmetrization.
The complex adjoint representation is irreducible, so Burnside's
theorem gives all complex color endomorphisms after complexification.
Since the words are real, their real span is
\(\operatorname{End}(\mathbb R^N)\). Thus on a single axis we may
take the bracket with \emph{arbitrary} real color matrices \(B_1,B_2\),
giving
\[
 [i\sigma_1\otimes B_1,i\sigma_1\otimes B_2]
       =-1\otimes[B_1,B_2].
\]
These commutators span \(\mathfrak{sl}(N,\mathbb R)\), which
contains all real skew matrices. Together with the imaginary
quaternionic axes times real symmetric matrices they include
\(\mathfrak{sp}(N)\).

The compact group \(\mathrm{Sp}(N)\) can rotate any occupied plane
onto any other. Both signs of every control occur as \(X\) ranges
over the Cartesian matrices, so finite products of constant-control
evolutions approximate such a rotation. Rescaling durations and
amplitudes fits the path into any prescribed positive time, with short
ramps joining it to the specified endpoints. On an open tube about the
resulting path the determinant stays positive and the scalar potential
is bounded. This tube has positive Brownian bridge measure. At fixed
time the transport norm is bounded by \(\exp(C_t\sup|X|)\), which is
integrable against the bridge. One can first use the heat kernel in
an enclosing Dirichlet ball and exhaust Cartesian space. The
Hamiltonian is nonnegative before projecting to singlets, since
tracing the supersymmetry anticommutator over spin removes the Gauss
term with its traceless Pauli coefficient. Integration against
\(e^{-st}\) now gives strictly positive Green pairings between
nonzero \(C\)-vectors at disjoint endpoints.

\paragraph{The Dirichlet current in the physical sector.}
The full-space propagation result applies to a solution in an
individual valley once we know the sign of its current. For compact
real physical \(f\) in that cone, the condition is encoded by the
equivalence
\begin{equation}
 \int_{D_i}(\phi_i(-s),f)=0
 \quad\Longleftrightarrow\quad
       e^{\gamma x}(H_{D_i}+s)^{-1}f\in L^2 .
 \label{tx:dualrange}
\end{equation}
The weighted inverse with free cap slope shows that
\(L_-=e^{-\gamma x}(H_{D_i}+s)e^{\gamma x}\) is onto and that its
kernel is spanned by \(e^{-\gamma x}\phi_i(-s)\). The two operators with
opposite exponential weights have domain \(D(H_{D_i})\).
The angular lower bound makes \(\partial_\rho\) infinitesimally
bounded relative to \(H_{D_i}\), so integration by parts and
invertibility at a sufficiently large shift identify the opposite
drift as \(L_-^*\). Its range is closed and orthogonal to
\(\ker L_-\). If the left side of \eqref{tx:dualrange} holds, solve
the adjoint equation with source \(e^{\gamma x}f\) and undo the
weight. Uniqueness gives the unweighted Dirichlet resolvent on the
right. Conversely, pair that resolvent with a radial cutoff of
\(\phi_i\) equal to one below \(R\) and zero above \(R+1\).
The error terms pair \(\phi_i\) or \(\partial_\rho\phi_i\) on this
shell with \((H_{D_i}+s)^{-1}f\). They tend to zero by weighted
Cauchy--Schwarz under the right side of \eqref{tx:dualrange}.

If two compact \(C^*\)-valued sources had opposite signs against
\(\phi_i\), a positive linear combination would pair to zero and
have a rapidly decaying Dirichlet response by \eqref{tx:dualrange}.
Each response is dominated in \(C^*\)-order by a fixed multiple
of their sum. The norm inequality for this order forces both
responses to have the same decay, contradicting their nonzero
pairings. Thus the functional has one sign on dual-positive
physical sources. Since \(\phi_i\) is invariant and symmetry
averaging preserves \(C^*\), the same sign holds on localized
tests before projection. By bipolarity \(\phi_i\) takes values
in \(C\) up to an overall sign. The diagram automorphism transports
the paired volumes and the ground mode between the valleys, so
their signs can be chosen consistently.

On either Dirichlet face the inward derivative lies in \(C\), as a
limit of one-sided difference quotients. The cap current is nonzero on
an open patch, since its projection against the angular ground mode
at the cap equals one. Returning to Cartesian density supplies a
positive factor. The zero extension \(\widetilde\phi_i\) has source
\(d_i^0=(H+s)\widetilde\phi_i=\tfrac12\partial_{\rm out}\phi_i\,d\sigma\).
All outward face currents have the same sign, and the two nonzero
cap patches have strictly positive Green pairing. Integrating over
the remaining faces adds a nonnegative term, so
\begin{equation}
 (d_j^0,(H+s)^{-1}d_i^0)>0 .
 \label{tx:strict}
\end{equation}
One may pair the currents by form duality or smooth them in disjoint
collars, using exponential decay along the sides to remove the
large-radius cutoff.

\subsection{Identification with the scattering amplitude}

Choose a smooth \(\chi_i\) inside each cone that vanishes near its
Dirichlet faces and is one at large radius on a smaller cone
containing the ground mode. The zero extension of \(\Phi_i=\chi_i\phi_i\)
has defect \(d_i(z)=(H-z)\Phi_i=[H,\chi_i]\phi_i\), supported near
the cap and in the side collar. The collar estimates give holomorphy
in \(L^2\) across the real energy interval and uniform bounds on
\(\|\langle X\rangle^b d_i(z)\|\) on compact \(z\)-sets for every
finite \(b\). Truncation in radius gives holomorphy in each weighted
space as well. The bilinear resolvent pairing \({\cal F}_{ji}\) of
\eqref{eq:analytic} is consequently holomorphic on the connected upper
side of the interval and across \(-s\). At \(-s\) the difference
\(\widetilde\phi_i-\Phi_i\) belongs to the energy form domain. Replacing
\(d_i(-s)\) and \(d_j(-s)\) by the two conormal sources leaves
\({\cal F}_{ji}(-s)\) unchanged, since each added term pairs disjoint
supports. Equation~\eqref{tx:strict} therefore gives
\({\cal F}_{ji}(-s)>0\).

\paragraph{Traveling waves on the radial half-line.}
Fix \(E=e_0+k^2/2\in I\), \(k>0\). Given a bounded scalar amplitude
\(a\) with \(\|a\|_{C^1}=M\), solve the perpendicular equation with
forcing \(-Ba\). A small truncated exponential weight centered on a
unit forcing shell preserves positivity whenever \(E+c^2/2<m\).
Summing the estimate over forcing shells gives, uniformly on \(I\),
\[
 \|q_\perp\|_{H^1([l,l+1])}
 \le CM\sum_{l'\ge \lfloor R_0\rfloor}
        \frac{e^{-c|l-l'|}}{l'}\le C'M/l .
\]
The return into the scalar equation consists of a term bounded by
\(CM\rho^{-3/2}\) and the derivative of
\(\langle p',q_\perp\rangle=O(M\rho^{-2})\). In the scalar sine
integral we integrate this derivative by parts using \(q_\perp(R_0)=0\).
The resulting correction to \(\sin(k(\rho-R_0))/k\) has operator
norm \(O_I(R_0^{-1/2})\) on bounded \(C^1\) amplitudes. Its fixed
point lies in the weighted space and agrees with the continued
solution. In particular its real ground amplitude satisfies
\begin{equation}
 \begin{split}
 a_i(\rho,E)&=A_{i,+}(E)e^{ik\rho}+A_{i,-}(E)e^{-ik\rho}+o(1),
                      \qquad \rho\longrightarrow\infty,\\
 A_{i,\pm}(E)&=\frac{\pm e^{\mp ikR_0}}{2ik}
                    +O_I(R_0^{-1/2}),\qquad
 A_{i,-}(E)=\overline{A_{i,+}(E)} .
 \end{split}
 \label{tx:travel}
\end{equation}
Reality gives the last identity, and increasing \(R_0\) keeps both
coefficients bounded away from zero on \(I\).

For \(b_i\in C_c^\infty(\operatorname{int}I)\) form the two packets
\[
 v_i(t)=\int e^{-itE}\Phi_i(E)b_i(E)\,dE,\qquad
 d_i(t)=\int e^{-itE}d_i(E)b_i(E)\,dE .
\]
At negative times the coefficient of \(e^{-ik\rho}\) gives the
incoming packet, and at positive times the coefficient of \(e^{ik\rho}\)
gives the outgoing packet. Passing to these coefficients requires norm
control throughout the moving shell. In the sine integral the difference
from the limiting coefficient is an integral over \(t'>\rho\) with
integrand bounded by \(C_I(t')^{-3/2}\). For each \(t'\), Plancherel
in \(k\) bounds the traveling packet after restriction to
\(L<\rho<t'\). Integration in \(t'\) gives exterior error at most
\(C_I L^{-1/2}\|b_i\|\). The perpendicular component and the
\(O_I(\rho^{-1})\) change to the physical radial channel have
square-integrable tails. On bounded shells all these remainders have
square-integrable kernels in energy and position. The corresponding
error operator is therefore compact, so its action on the weakly
dephasing packet tends to zero in norm.

In the energy normalization \(\sqrt{2/(\pi k)}\sin(kr)\) the
incoming and outgoing coefficients of \(v_i(t)\) are
\[
 b_i^{\rm in}=-i\sqrt{2\pi k}\, A_{i,-}b_i,\qquad
 b_i^{\rm out}= i\sqrt{2\pi k}\, A_{i,+}b_i .
\]
Two integrations by parts in energy give \(d_i\in L^1(dt;L^2)\).
Thus \(e^{itH}v_i(t)\) has strong limits at both time signs. The
traveling-wave normalization identifies these limits with the two
isometric channel maps on their respective energy envelopes, which
are dense since the multipliers in \eqref{tx:travel} are bounded
away from zero. Writing \(W_{\pm,i}\) for the respective maps on
valley \(i\), the contraction \(S_{ji}=(W_{+,j})^* W_{-,i}\)
measures the selected matrix element even when other channels are open.

The exact incident solution is obtained by subtracting the retarded
response to the defect,
\[
 u_i^-(t)=v_i(t)-i\int_{-\infty}^t e^{-i(t-t')H}d_i(t')\,dt' .
\]
For the Hermitian inner product linear in the second argument,
\(\partial_t\langle v_j,u_i^-\rangle=-i\langle d_j,u_i^-\rangle\).
The disjoint cones give zero incoming overlap and
\(\langle d_j(t),v_i(t)\rangle=0\) for \(j\ne i\). Integrating in
time and using the retarded transform
\(\int_0^\infty e^{-it(H-E)-\epsilon t}\,dt
=-i(H-E-i\epsilon)^{-1}\) therefore gives
\begin{equation}
 \langle b_j^{\rm out},S_{ji}b_i^{\rm in}\rangle
 =2\pi i\lim_{\epsilon\downarrow0}\int
 \overline{b_j(E)}b_i(E)\,
 \langle d_j(E),(H-E-i\epsilon)^{-1}d_i(E)\rangle\,dE .
 \label{tx:retarded}
\end{equation}
The two time-integrable source packets justify removal of the damping.

\paragraph{The upper boundary value.}
The conjugated outgoing coefficient in the bra and the incoming
coefficient in the ket contribute
\(-2\pi k\,\overline{A_{j,+}}A_{i,-}=-2\pi k A_{j,-}A_{i,-}\).
To identify the right side of \eqref{tx:retarded} with the holomorphic
pairing, replace its bra by \(d_j(E-i\epsilon)\) and its ket by
\(d_i(E+i\epsilon)\). Their difference quotients converge in norm
uniformly on compact subintervals. By the spectral theorem,
\(\epsilon(H-E-i\epsilon)^{-1}\) and its adjoint have norm at most
one and converge strongly to zero off the at most countable
eigenvalues. Dominated convergence makes the integrated error tend to
zero. Schwarz reflection for the real solution gives
\(\mathcal J d_j(E-i\epsilon)=d_j(E+i\epsilon)\), as required for
the bilinear extension of the real pairing. Dividing the product of
normalizations by \(2\pi i\) in \eqref{tx:retarded} gives
\eqref{eq:boundary} first in distributions. Its right side is locally
bounded because \(S_{ji}(E)\) is contractive.

Boundary uniqueness needs the nontangential limit at individual
energies. Let \(P_H\) be the spectral resolution in the physical
sector. Around a real \(c\in\operatorname{int}I\), expand \(d_j(z)\)
and \(d_i(z)\) in a disk with real Taylor coefficients \(u_a,v_b\)
respectively. On \(K=[c-r,c+r]\) strictly within this disk define
the real signed measure
\[
 d\mu(\lambda)=\sum_{a,b\ge0}(\lambda-c)^{a+b}
       (u_a,P_H(d\lambda)v_b)_{\rm bil}.
\]
The series converges in total variation by the Taylor norm bounds.
For \(z\) near the interior of \(K\) the spectral theorem gives
\[
 {\cal F}_{ji}(z)=h(z)+\int_K\frac{d\mu(\lambda)}{\lambda-z},
\]
where \(h\) extends holomorphically through the interval and is real
there. The spectral integral outside \(K\) has both properties.
Within \(K\) the factor
\((z-c)^{a+b}-(\lambda-c)^{a+b}\) cancels the resolvent denominator.
The divided polynomials are locally summable by the same Taylor
bounds. The imaginary part of the upper distributional trace is
consequently \(\pi\mu\) on the interior. By \eqref{eq:boundary} the
measure there has a locally bounded density. After a smooth cutoff
inside \(K\) this density is in \(L^2\), so the Poisson extensions
of the density and its Hilbert transform have nontangential limits
equal to their distributional traces almost everywhere. The remaining
part of the measure is supported away from any smaller interval where
the cutoff equals one. Its Cauchy integral is analytic across that
interval, proving that \eqref{eq:boundary} holds nontangentially
almost everywhere on \(I\).

If \(S_{ji}(E)\) vanished on a set of positive measure, some upper
half-disk would have vanishing boundary value on a positive-length
subset of its diameter. The Luzin--Privalov uniqueness
theorem~\cite{Privalov} would then force \({\cal F}_{ji}\) to vanish
on the connected continuation, contradicting \({\cal F}_{ji}(-s)>0\).
The diagram involution exchanges the two directions of transmission.
A countable exhaustion by compact \(I\) proves nonvanishing almost
everywhere in the open band. The cap radius may increase as \(I\)
approaches either endpoint.


\begin{thebibliography}{99}
\addcontentsline{toc}{section}{\refname}
\raggedright
\bibitem{dWLN} B. de Wit, M. L\"uscher and H. Nicolai, \emph{The supermembrane is unstable}, \href{https://doi.org/10.1016/0550-3213(89)90214-9}{Nucl. Phys. B \textbf{320} (1989) 135--159}.
\bibitem{Michishita} Y. Michishita, \emph{On the relationship between discrete and continuous energy spectra of \(SU(N)\) supermembrane matrix model}, \href{https://doi.org/10.1016/j.nuclphysb.2013.09.011}{Nucl. Phys. B \textbf{877} (2013) 124--151}, \arref{1308.2786}.
\bibitem{Qi} M. Qi, \emph{The Matrix in Pieces: Complete BFSS Scattering at Finite Rank}, \arref{2611.31649}.
\bibitem{HW} M. Henningson and N. Wyllard, \emph{Low-energy spectrum of \(\mathcal N=4\) super-Yang--Mills on \(T^3\): flat connections, bound states at threshold, and \(S\)-duality}, \href{https://doi.org/10.1088/1126-6708/2007/06/001}{JHEP \textbf{06} (2007) 001}, \arref{hep-th/0703172}.
\bibitem{Smilga} A.V. Smilga, \emph{Born--Oppenheimer corrections to the effective zero-mode Hamiltonian in SYM theory}, \href{https://doi.org/10.1088/1126-6708/2002/04/054}{JHEP \textbf{04} (2002) 054}, \arref{hep-th/0201048}.
\bibitem{LeeYi} S.-J. Lee and P. Yi, \emph{Witten index for noncompact dynamics}, \href{https://doi.org/10.1007/JHEP06(2016)089}{JHEP \textbf{06} (2016) 089}, \arref{1602.03530}.
\bibitem{Ambrozinski} Z. Ambrozi\'nski, \emph{\(D=4\) supersymmetric Yang--Mills quantum mechanics with three colors}, \arref{1409.7435}.
\bibitem{PST} C. Pedder, J. Sonner and D. Tong, \emph{The Berry Phase of D0-Branes}, \href{https://doi.org/10.1088/1126-6708/2008/03/065}{JHEP \textbf{03} (2008) 065}, \arref{0801.1813}.
\bibitem{BRY} B. Balthazar, V.A. Rodriguez and X. Yin, \emph{Hamiltonian truncation study of supersymmetric quantum mechanics: S-matrix and metastable states}, \href{https://doi.org/10.1007/JHEP08(2019)100}{JHEP \textbf{08} (2019) 100}, \arref{1610.07275}.
\bibitem{Ito} K. Ito and E. Skibsted, \emph{Stationary scattering theory on manifolds}, \href{https://doi.org/10.5802/aif.3417}{Ann. Inst. Fourier \textbf{71} (2021) 1065--1119}.
\bibitem{BMS} M. Braverman, O. Milatovic and M. Shubin, \emph{Essential self-adjointness of Schr\"odinger-type operators on manifolds}, \href{https://doi.org/10.1070/RM2002v057n04ABEH000532}{Russ. Math. Surv. \textbf{57} (2002) 641--692}, \arref{math/0201231}.
\bibitem{Luscher} M. L\"uscher, \emph{Some analytic results concerning the mass spectrum of Yang--Mills gauge theories on a torus}, \href{https://doi.org/10.1016/0550-3213(83)90436-4}{Nucl. Phys. B \textbf{219} (1983) 233--261}.
\bibitem{Wosiek} J. Wosiek, \emph{Spectra of supersymmetric Yang--Mills quantum mechanics}, \href{https://doi.org/10.1016/S0550-3213(02)00810-6}{Nucl. Phys. B \textbf{644} (2002) 85--112}, \arref{hep-th/0203116}.
\bibitem{Knapp} A.W. Knapp, \emph{Lie Groups Beyond an Introduction}, second ed., Birkh\"auser, Boston (2002), Corollary~4.51.
\bibitem{Privalov} N. Lusin and J. Priwaloff, \emph{Sur l'unicit\'e et la multiplicit\'e des fonctions analytiques}, \href{https://doi.org/10.24033/asens.763}{Ann. Sci. \'Ecole Norm. Sup. \textbf{42} (1925) 143--191}.
\bibitem{BajajLin} K. Bajaj and H.W. Lin, \emph{Solving the \(SU(2)\) BFSS model}, \arref{2611.14512}.
\bibitem{GoleniaJecko} S. Gol\'enia and T. Jecko, \emph{A new look at Mourre's commutator theory}, \href{https://doi.org/10.1007/s11785-007-0011-4}{Complex Anal. Oper. Theory \textbf{1} (2007) 399--422}, \arref{math/0607275}.
\end{thebibliography}
\end{document}